# Enhancing 5G NTN VSAT RACH in GNSS-Denied Environments: A Multi-Step Single-Satellite AoA/FoA/ToA Positioning Using Levenberg–Marquardt Kalman Filtering

**Francesco Menzione[1], Alejandro Gonzalez-Garrido[1,], Flavien Ronteix-Jacquet[1] and Ottavio M. Picchi[2]**

[1] European Commission, Joint Research Centre; name.surname@ec.europa.eu
[2] External consultant for European Commission, Joint Research Centre; ottavio.picchi@ext.ec.europa.eu
* Correspondence: francesco.menzione@ec.europa.eu.

**Abstract**

The 3GPP 5G Non-Terrestrial Networks (NTN) technology fundamentally relies on a Global Navigation Satellite System (GNSS) fix at the User Equipment (UE) to pre-compensate for user-link delay and Doppler shifts prior to any transmission, including the random-access procedure for the initial access. In GNSS-denied or interfered environments, this dependency triggers preamble collisions and connection dropouts, undermining satellite-based global connectivity. To address this vulnerability, this paper introduces a novel single-satellite positioning framework designed for higher-frequency (Ku/Ka-band) 5G NTN Very Small Aperture Terminal (VSAT) equipped with a phased-array antenna. The multi-step scheme derives an initial coarse position from Angle of Arrival (AoA) observables, extracts Frequency of Arrival (FoA) and Time of Arrival (ToA) measurements from Low Earth Orbit (LEO) downlink reference signals and hybridizes them within a Levenberg–Marquardt navigation filter. Systematic evaluations across static grids and realistic dynamic urban trajectories detail the incremental gains of transitioning from single-epoch AoA error minimization ($J_0$) to multi-epoch angular tracking ($J_1$), followed by the integration of Doppler ($J_2$) and ranging observables ($J_3$). Grounded in a realistic error budget, the proposed framework reduces initial positioning uncertainty from tens of kilometres down to a few kilometres. This level of positioning accuracy combined with the GNSS-resilient features introduced in 3GPP Release 20 NR NTN specifications, enable robust initial access in compromised operational environments with minimal relaxation of 3GPP standards.

**Keywords**: 5G NR-NTN; PRACH; initial random access; GNSS-denied positioning; Angle of Arrival (AoA); Frequency of Arrival (FoA), Time of Arrival (ToA), Levenberg-Marquardt Kalman Filter; Ka-band VSAT; Low Earth Orbit constellations

# 1 Introduction

In Release 17 of the 3GPP 5G NR standard, support for NTN was introduced for space-based communication. In the architecture studied and adopted for NR-NTN [1], the UE has to perform all necessary timing and frequency corrections prior to any uplink transmission to avoid misalignment at the reference point (e.g., at the satellite). To achieve this, the UE must acquire a valid GNSS position alongside broadcasted network data, including satellite ephemerides and common timing advance, to pre-compensate for uplink timing, frequency offset, and Doppler shift prior to transmitting Message 1 (Msg1) over the Physical Random Access Channel (PRACH). The standard [2] defines a set of timing error and frequency offset requirements for both initial access and connected mode. These requirements are typically met using a GNSS receiver, which provides the UE with its location and accurate absolute-time reference. In some scenarios, however, GNSS availability cannot be taken for granted. For example, [3] reports the times and locations of GNSS jamming, demonstrating a rise in jamming events, especially in conflict zones. Furthermore, GNSS jamming is one of the easiest ways to disrupt communication services, as demonstrated by the Starlink disruptions in Iran [4]. While a key motivation for adopting NTN is providing reliable coverage and robust connectivity globally, the current design remains heavily dependent on GNSS availability.

To address GNSS vulnerability, 3GPP conducted a feasibility study on GNSS-resilient NTN operations across diverse satellite orbits (NGSO/GSO), frequency bands (FR1/FR2), and terminal types (handheld/VSAT) under degraded GNSS conditions [5]. The study grouped initial-access solutions into three main categories [6]: i) reducing UE location uncertainty via downlink reference signals (e.g., exploiting Downlink Time Difference of Arrival); ii) expanding gNB timing advance (TA) and frequency offset tolerances; and iii) utilizing network-provided assistance data (e.g., cell reference location, common TA, and Doppler). Technical contributions explored different approaches [7], [8], [9], each presenting operational trade-offs.

Focusing on the first category, uplink TA estimation and carrier frequency offset (CFO) pre-compensation for GNSS-deprived UEs have also been addressed in the scientific literature. To solve the PRACH challenge without GNSS, [10] proposes a two-step approach: an iterative least-squares algorithm processing multi-satellite SSB Doppler shifts to estimate position and CFO, combined with a CP-free preamble design utilizing flexible sequence cascading and differential power allocation to enhance detection.

Most methods proposed in 3GPP and in scientific literature to drastically reduce location uncertainty rely on receiving signals from multiple satellites simultaneously, a feature not currently supported in existing specifications, at least for terminals operating at higher frequencies (Ku/Ka-band). In these bands, current specifications contemplate initial access through a single serving satellite at a time, where a VSAT is necessary to close the link budget.

On one hand, a VSAT terminal equipped with a phased-array antenna offers the key advantage of deriving AoA measurements by steering a single high-gain receive beam. Exploiting the corresponding AoA measurement to perform coarse positioning is a widely investigated solution [11], [12], [13]. However, those solutions are generally limited to accuracies on the order of dozens of kilometres , whereas providing reliable PRACH access requires moving toward higher accuracy [14].

To address this limitation, this paper proposes a novel integrated positioning framework that combines the standard AoA approach with additional receiver capabilities adapted from Signals of Opportunity (SoP) navigation [15],[16]. Although the technology required to derive additional ToA and FoA observables from standard synchronization patterns, such as the Primary Synchronization Signal (PSS) and Secondary Synchronization Signal (SSS), is already consolidated [17] and proposed for 3GPP signals in [18],[19],[10], the specific impact of this processing on single-satellite VSAT terminal localization for improved PRACH needs further investigation. To move forward in accuracy, we propose a multi-step framework where an initial coarse AoA positioning fix enables the proper initialization and reconstruction of the Doppler (FoA) and ranging (ToA) observables. Building upon this multi-observable reconstruction, our positioning engine integrates the measurements within a robust multi-epoch processing scheme based on a custom Levenberg–Marquardt Kalman Filter (LMKF). This scheme drives significant gains in positioning accuracy by exploiting both the time-correlation of user positions and the geometric evolution across sequential single-satellite passes.

The paper validates this approach, without loss of generality, against a representative 5G NTN scenario, accounting for system capabilities as well as AoA, FoA, and ToA measurement characterization derived from expected signal characteristics, signal propagation impairments, and VSAT user terminal parameters.

The assessment includes an analysis of the incremental contributions of the different observables, transitioning from standard single epoch to multi-epoch angular tracking, followed by the integration of Doppler and ranging observables. Finally, the analysis evaluates the resulting location uncertainty and its direct impact on timing advance during the RACH procedure, considering the geometry of the RACH and suggesting possible way forwards to optimize this stage.

This framework demonstrates that VSAT-accessible observables alone enable kilometer-level positioning. Crucially, it is compatible with GNSS-resilient features, such as the two-sequence PRACH transmission introduced in Rel-20, making positioning more robust. However, a minimal relaxation of radio resource management (RRM) requirements is still necessary to ensure successful PRACH preamble reception at the satellite side under GNSS-denied conditions.

The remainder of this paper is structured as follows. Section 2 describes the 5G-NTN target scenario, encompassing the space segment, the 5G signal model, and VSAT assumptions. Section 3 provides the full error budget when generating positioning observables, including signal processing and propagation impairments. Section 4 introduces the multi-step procedure and the LMKF core algorithm. Section 5 discusses the simulation results. Finally, Section 6 concludes the paper.

## 2 5G-NTN system and UE assumptions

This subsection outlines the system assumptions adopted in this paper to evaluate the proposed method under realistic 5G-NTN scenarios.

*A. Space Segment*

The first assumption relates to the satellite constellation, which is based on an inclined walker-star shell and a polar one both at an altitude of 1200 km. The orbital parameters for the constellation used in this study are listed in Table 1, while a visual representation of the constellation is shown in Figure 1.

Furthermore, we consider a satellite footprint defined by the parameters of Set-2 in [20] for LEO-1200 km, specific to Ka-band. Notably, a polar shell was added to the inclined shell proposed in [22] to reduce location uncertainty at higher latitudes, which the inclined shell alone cannot adequately cover. Additionally, we consider minimum elevation angle (MEA) of 20° elevation according to [21].

**Table 1.** Orbital parameters for the simulated LEO-1200 km constellation

| Parameter | Inclined shell | Polar shell |
|---|---|---|
| Satellite altitude | 1200 km | 1200 km |
| Semi major axis | 7578.17 km | 7578.17 km |
| n. of satellites | 264 | 45 |
| n. of orbital planes | 22 | 3 |
| n. of satellites per plane | 12 | 15 |
| Plane inclinations | 55° | 85° |
| Intra-plane phasing | 30° | 24° |
| Inter-plane phasing | 16.36° | 30° |

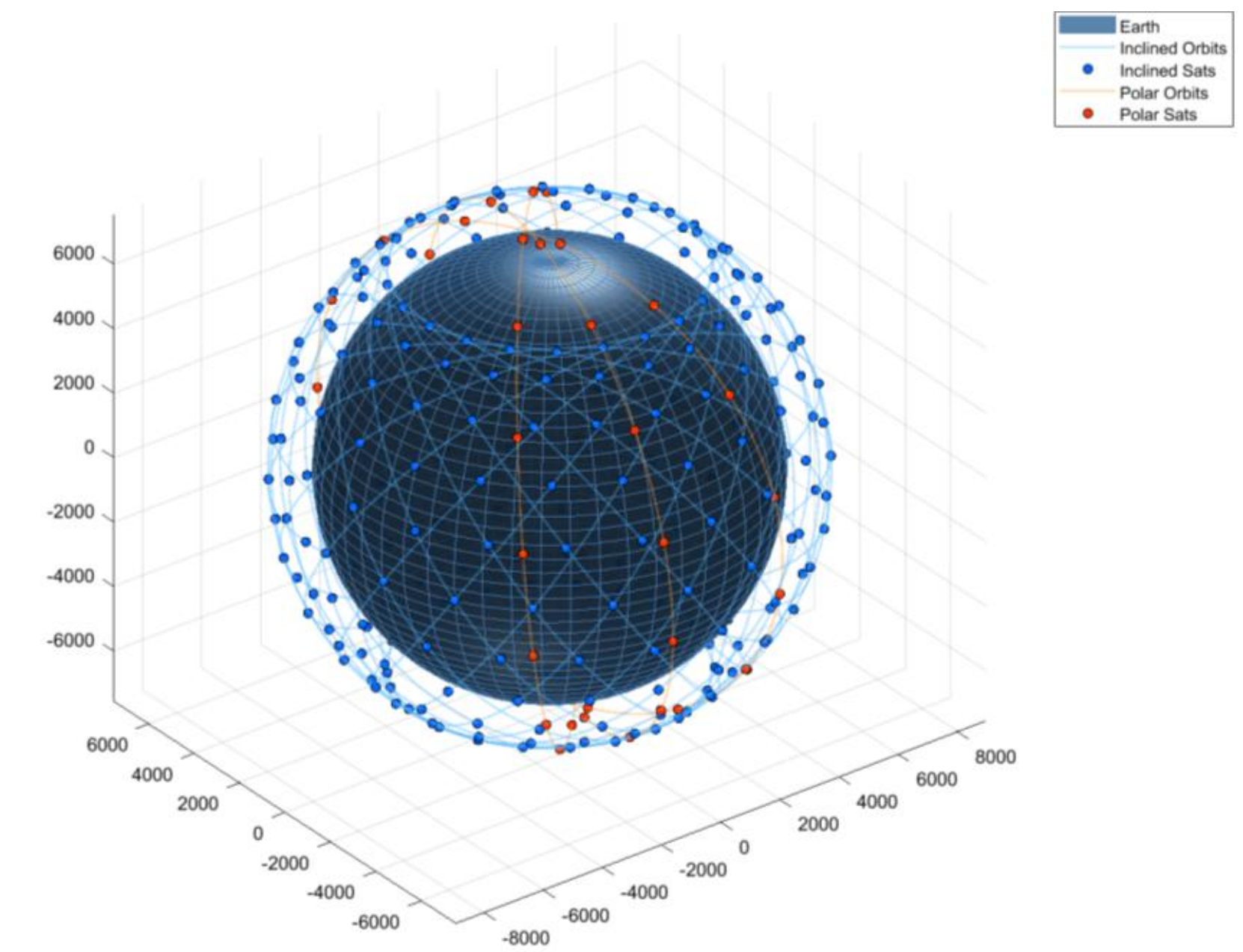

**Figure 1.** Appearance of the LEO-1200 km constellation.

The satellite footprint visualization generated by our simulator is shown in Figure 2, as already presented in [19], [23]. Figure 2 shows the satellite coverage for a LEO satellite, whose sub-satellite point falls at latitude, longitude, altitude (LLA) coordinates (0,0,0). It is worth noting that beams are arranged in rings and a greater number of rings corresponds to a larger footprint. Blue beams represent beams between ring 1 and ring 12 (elevation mask up to 30°), red beams are the beams in the 13$^{th}$ ring (elevation mask up to 25°), whereas green beams are those in the 14$^{th}$ ring (elevation mask up to 20°). Figure 2 indicates that the minimum number of rings required to cover the region down to a 20° elevation is fourteen, corresponding to 547 beams. Figure 2 presents the satellite antenna parameters used in our simulator [19]. Our link budget parameters match those in [22] exactly.

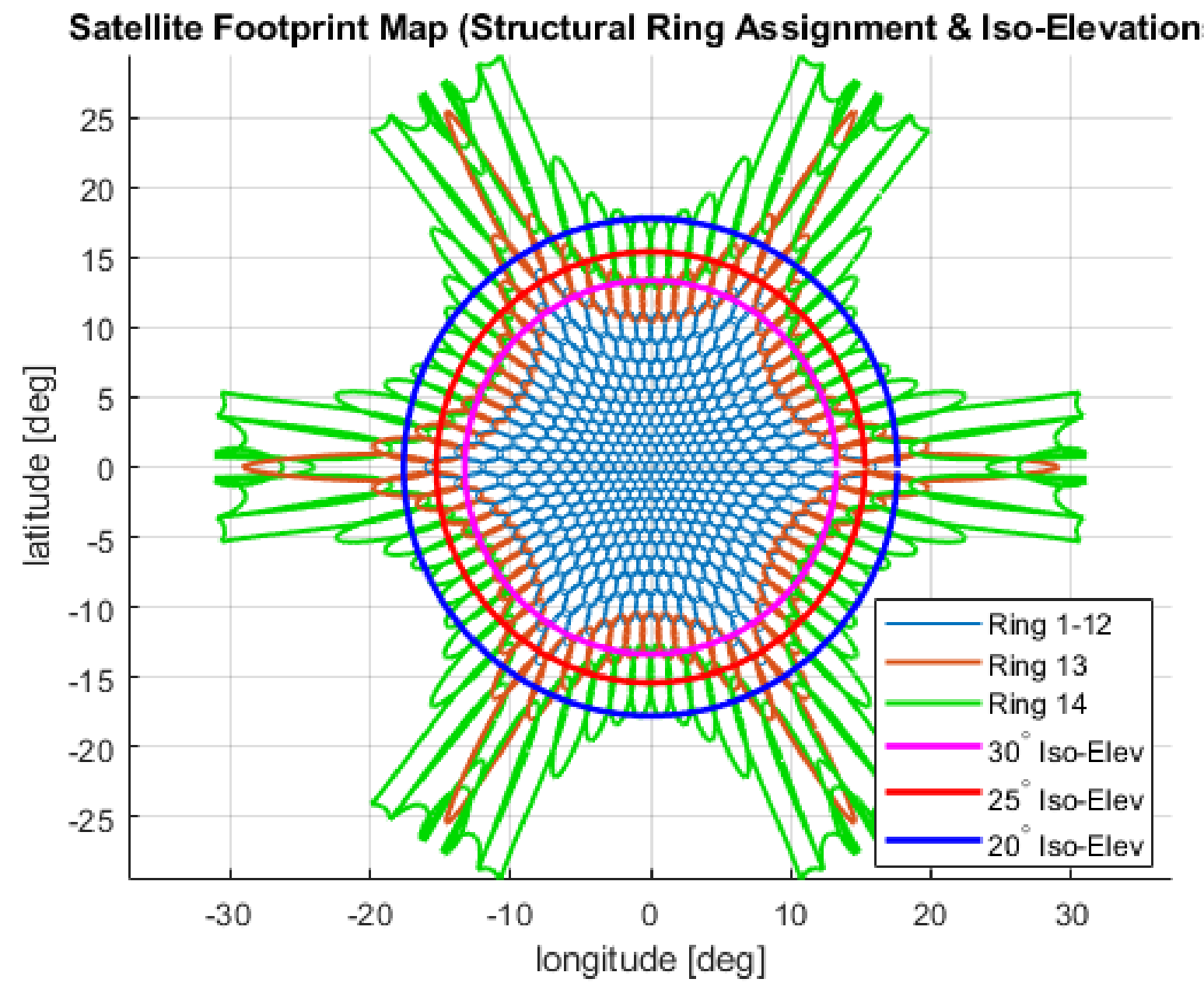

**Figure 2.** Satellite footprint using antenna parameters proposed in [20].

**Table 2.** Satellite antenna parameters.

| Parameter | Value |
|---|---|
| Equivalent Sat. Ant. Aperture | 0.2 m |
| Satellite Tx max gain | 30.5 dB |
| HPBW | 4.4127° |
| Beam diameter D (Nadir) | 90 km |
| Adj. Beam Spacing (UV plane) | 0.0667 |

It is worth recalling that the VSAT is intended to receive the downlink signal from only one satellite at a time. To obtain simulation windows that are longer than the typical visibility period of a single LEO satellite (usually a few minutes), the simulator has been configured to emulate satellite handover. Consequently, the satellite that is observed from the user with the highest elevation is selected as the serving satellite for the corresponding ground cell. This satellite remains active over the cell until it falls below a prescribed elevation threshold. The threshold does not necessarily coincide with the MEA, because it may be defined by the scenario under investigation; for example, an urban environment could require a more stringent elevation limit than that imposed by the satellite footprint.

Figure 3 illustrates the temporal evolution of the serving satellite's elevation angle and C/N0 under the selected handover policy. Notably, a satellite handover is emulated as soon as the serving satellite reaches a minimum elevation threshold of 40°. This threshold should not be confused with the masking angle; rather, it serves as a straightforward mechanism to emulate a ground-segment scheduling policy. When a handover occurs, a discrete jump appears in both the elevation and C/N0 curves. Maintaining this 40° minimum elevation ensures that downlink signals are received with a high probability of an unobstructed line-of-sight, even in dense urban environments.

It is worth noting that the ripple appearing in the bottom sub-plot in Figure 3 is related to the beam changes due to the Earth moving beam assumption. This ripple is typically 3 dB, corresponding to the half-power beam width (HPBW) of each satellite beam.

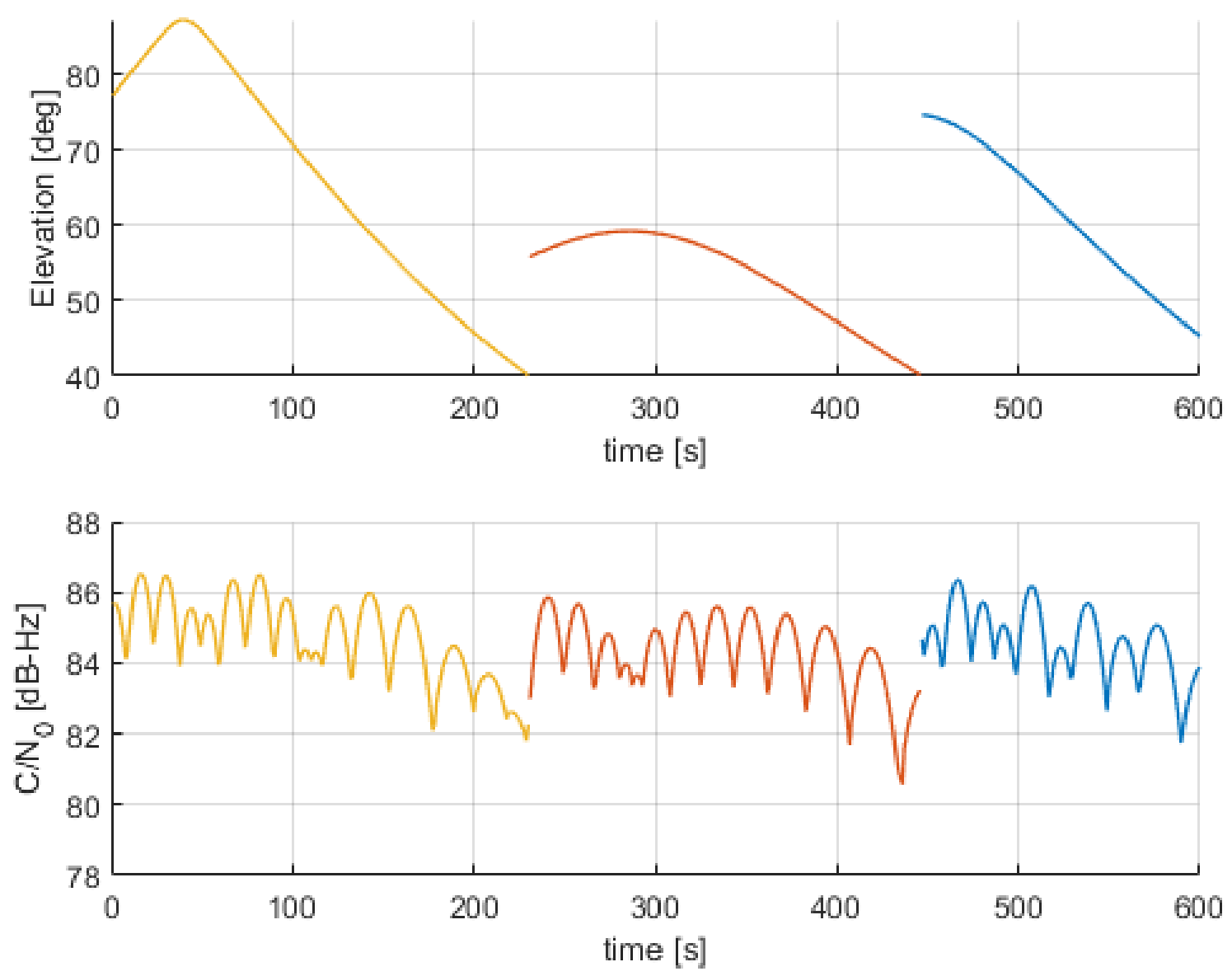

**Figure 3.** Serving satellite elevation and C/N0 over time.

*B. 5G signal model*

In this work, we assume the downlink physical layer to be built upon orthogonal frequency division multiplexing with a cyclic prefix (CP-OFDM), used in 5G NR. The core component responsible for initial cell search, synchronization, and beam management is the synchronization signal block (SSB) within the physical broadcast channel (PBCH). The time-domain baseband signal for a 5G NR downlink slot is generated by modulating the mapped complex data symbols onto the OFDM subcarriers. For a given subcarrier spacing configuration $\mu$, the continuous-time signal can be modeled as

$$s^{\mu}(t) = \sum_{l=0}^{N_{symb}^{slot}-1} \sum_{k=0}^{N_{RB}^{\mu} N_{SC}^{RB}-1} a_{k,l} \cdot e^{j2\pi k \Delta f (t - N_{CP,l}^{\mu} T_c - l T_{symb}^{\mu})} \operatorname{rect}\left(t - T_{symb}^{\mu}\right) \tag{1}$$

where the parameters are defined comprehensively as:

- $a_{k,l}$ represents the complex-valued modulation symbol mapped to Resource Element (RE) at subcarrier index $k$ and OFDM symbol index $l$.
- $\Delta f$ is the subcarrier spacing, where $\Delta f = 15 \times 2^{\mu}$ kHz.

- $N_{RB}^{\mu}$ is the maximum transmission bandwidth configuration for the given numerology.
- $N_{sc}^{RB}$ is the number of subcarriers per resource block (RB), and it is fixed at 12.
- $N_{CP,l}^{\mu}$ is the cyclic prefix length factor for symbol $l$.
- $T_c$ is the basic 5G NR time unit defined as $1/(\Delta f_{max} \cdot N_f)$, resulting around 0.509 ns.
- $T_{symb}^{\mu}$ is the total duration of an OFDM symbol including its CP length.
- $\mathrm{rect}(t - T_{symb}^{\mu})$ is a rectangular window of length $T_{symb}^{\mu}$.

The SSB is a localized cluster of 4 consecutive OFDM symbols and 240 contiguous subcarriers (20 RB). It contains three signals: i) the primary synchronization signal (PSS); ii) the secondary synchronization signal (SSS); iii) the PBCH. In this work we assume a subcarrier spacing (SCS) of 120 kHz, i.e. the smaller option for a signal operating in FR2 [38]. With this high spacing, the duration of a single OFDM symbol is very short, approximately $\frac{1}{120\ kHz} \approx 8.33\mu s$. A 1 ms subframe therefore contains eight slots, giving a slot duration of only 0.125 ms. In this work, we consider only the PSS and SSS. Both the PSS and SSS occupy only the central 127 subcarriers. Consequently, the bandwidth occupied by the PSS/SSS is $127 \times 120\ kHz = 15.24\ MHz$.

*C. VSAT User Equipment assumptions*

While this work primarily addresses VSAT terminals operating in the high frequency FR2 regime (specifically Ka-band), the methodologies and results presented remain broadly applicable to lower frequency bands, such as Ku-band. In all cases, the user equipment (UE) is assumed to be a VSAT terminal equipped with a phased-array antenna, a structural requirement necessary for Angle of Arrival (AoA) estimation. The VSAT is assumed to be capable of steering one single receiving beam toward the serving satellite. A gain of approximately 40 dBi is assumed for the VSAT, in line with [20] and [22]. Consequently, the number of array elements required to achieve this gain exceeds 4000, assuming each element provides a maximum gain of 3 dB. The resulting receiving aperture for the Ka band (20 GHz) is roughly 50 × 50 cm. Although complex, this antenna design closely aligns with commercial products such as [45], which features a G/T of 13 dB/K, corresponding to 38 dBi, assuming a system temperature of 300K. The product in [45] operates in the receive band within 17.7 and 21.2 GHz and in the transmit band within 27.5 and 31.0 GHz. Despite being designed for military applications its adaptation to 5G NTN waveforms appears feasible, demonstrating the validity of our UE antenna assumptions.

To minimize overall cost and power consumption compared to digital beamforming architectures, an analogue or hybrid beamforming approach can be assumed [24]. It is worth noting that the VSAT is expected to blindly acquire the AoA by scanning the entire sky region, and subsequently track the AoA of the serving satellite. For analog beamforming, this acquisition stage may require a serial search across a large set of possible candidate angles. In contrast, hybrid architectures can accelerate the acquisition using a reduced MUSIC algorithm or Bayesian compressive sensing techniques [25], [26]. The practical feasibility of blind AoA detection in commercial terminals is demonstrated by solutions such as [45], which natively support this feature. However, given its relevance, the topic of blind AoA acquisition deserves dedicated study and is beyond the scope of this paper.

As outlined in Section 1, this paper investigates the opportunistic leverage of additional ToA and FoA measurements. To this end, the UE is assumed to be equipped with a dedicated Software-Defined Radio (SDR) module, consistent with the experimental framework presented in [19]. This module enables the acquisition, tracking, and extraction of Doppler and code-phase observables directly from single-satellite PSS and SSS components. This architecture builds upon findings from our previous studies [17], [18], which demonstrated both the feasibility and seamless integration of such an SDR front-end into next-generation VSAT user terminals. While various hardware architectures can execute this front-end processing, our primary focus is on the downstream navigation engine: specifically, how these extracted AoA, FoA, and ToA observables are reconstructed, managed, and exploited. A comprehensive analysis of the expected performance and full error budget associated with this VSAT configuration is presented in the following section.

# 3 AoA, FoA and ToA observable generation in presence of signal propagation and processing impairments

Figure 4 illustrates the overall architecture of the simulation environment developed for this study, which serves as the primary reference throughout the rest of the paper. It evaluates the positioning algorithm's performance using realistic AoA, FoA, and ToA observables generated according to the system and user terminal assumptions introduced in the previous section. For completeness, a detailed mathematical formulation of these observables is provided in Appendix A and the notation and terminology throughout the remainder of this section adopt those defined in this appendix. The following subsections detail the expected Cramér-Rao Lower Bounds (CRLBs) for the receiver-side estimations, alongside the additional impairments affecting the observables and expected to affect a 5G NTN positioning system operating in the considered Ka band.

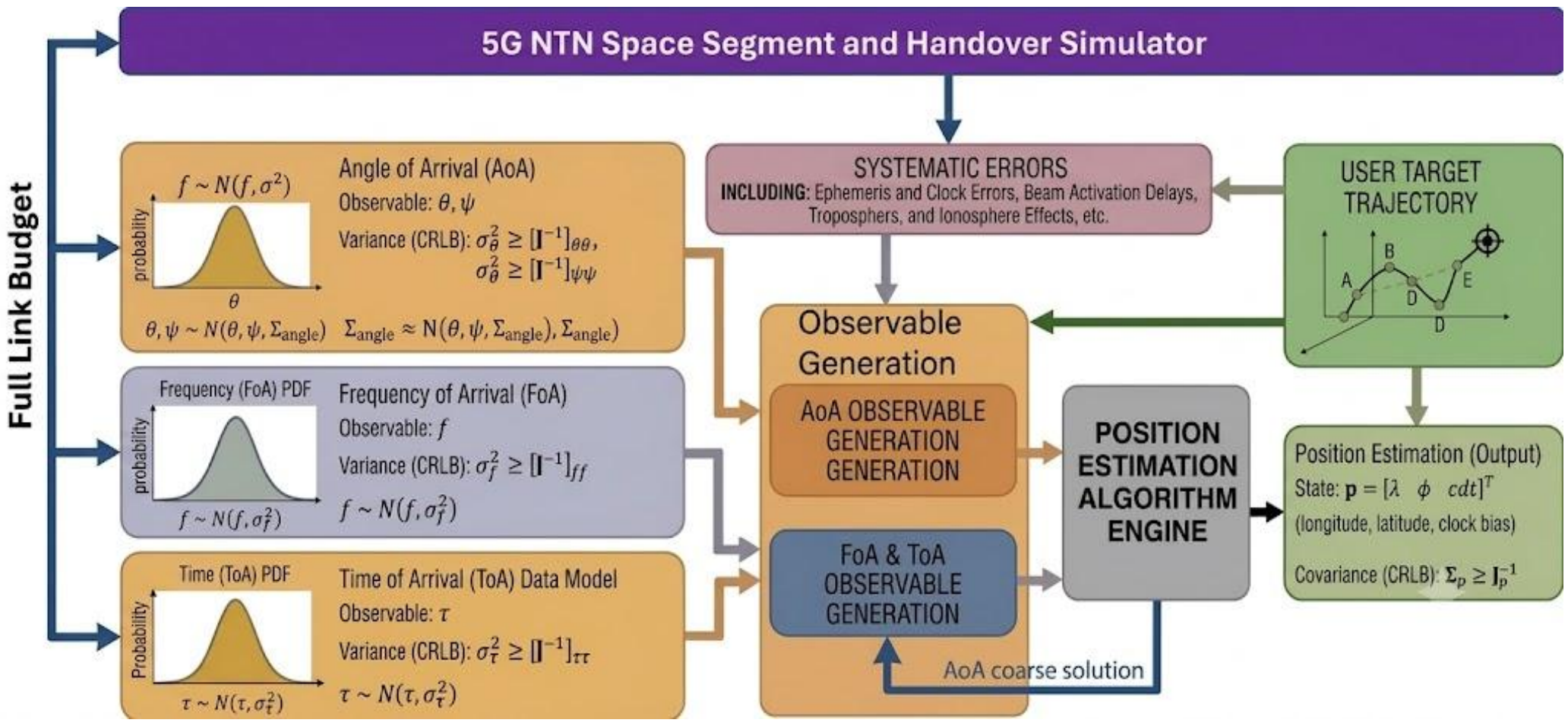

**Figure 4.** Error models considered in our simulator.

## 3.1 *Expected receiver processing error*

Figure 5 shows the CRLB for the estimation of azimuth, elevation, range rate, and range estimation from 5G-NTN VSAT signal processing.

*A. Angle Of Arrival*

Let us examine the first two subplots in Figure 5, which are specific to the AoA (azimuth and elevation). For these two CRLBs, the framework established in the work [25] has been followed, where the CRLB takes the following expression:

$$\mathbf{C} = \frac{\sigma^2}{2L \cdot P}\{\mathrm{Re}[\mathbf{D}^H \mathbf{P}_{\mathbf{A}}^{\perp} \mathbf{D}]\}^{-1} \quad (2)$$

where matrix $\mathbf{C} \in \mathbb{R}^{2\times 2}$ and it is the inverse of the Fisher information matrix (FIM) for the joint estimation of azimuth $\phi$ and elevation $\theta$. The CRLBs for $\phi, \theta$ are simply the diagonal elements of $\mathbf{C}$. The term $L$ indicates the number of snapshots, $\sigma^2$ represents the variance of the thermal noise and $P$ denotes the useful signal power, measured at each radiating element. Meanwhile, $\mathbf{P}_{\mathbf{A}}^{\perp} = \boldsymbol{I} - \frac{\boldsymbol{a}\boldsymbol{a}^H}{N_x N_y}$ is the orthogonal projection matrix that projects the columns of the derivative matrix onto the noise subspace, which is perpendicular to the steering vector $\boldsymbol{a}$. The terms $N_x$ and $N_y$ represent the number of array elements, distributed along the $x$-axis and $y$-axis, respectively, assuming that the UT is equipped with a planar array. Finally, the spatial derivative matrix $\mathbf{D} = \left[\frac{\partial \boldsymbol{a}}{\partial \boldsymbol{\theta}} \; \frac{\partial \boldsymbol{a}}{\partial \boldsymbol{\phi}}\right]$ tracks how the steering vector phase shifts across both angular coordinates. The first two subplots (the top graphs) in Figure 5 have been obtained by assuming a square array with 64x64 elements. This number of elements, yielding an array factor of around 36 dB, is compatible with a maximum receiver gain of 40 dB required for the link budget closure in [20]. It is worth noting that these CRLBs are plotted against the element SNR, which corresponds to the SNR measured at each individual antenna element, i.e. without accounting the array factor. So looking at these subplots, considering an array SNR of 2 dB (in line some study cases in [20]), the element SNR turns out to be around -34 dB. This element SNR provides an error $\upsilon_{\phi}$ of 0.2° on the azimuth estimation and an error $\upsilon_{\theta}$ of 0.07° on the elevation estimation.

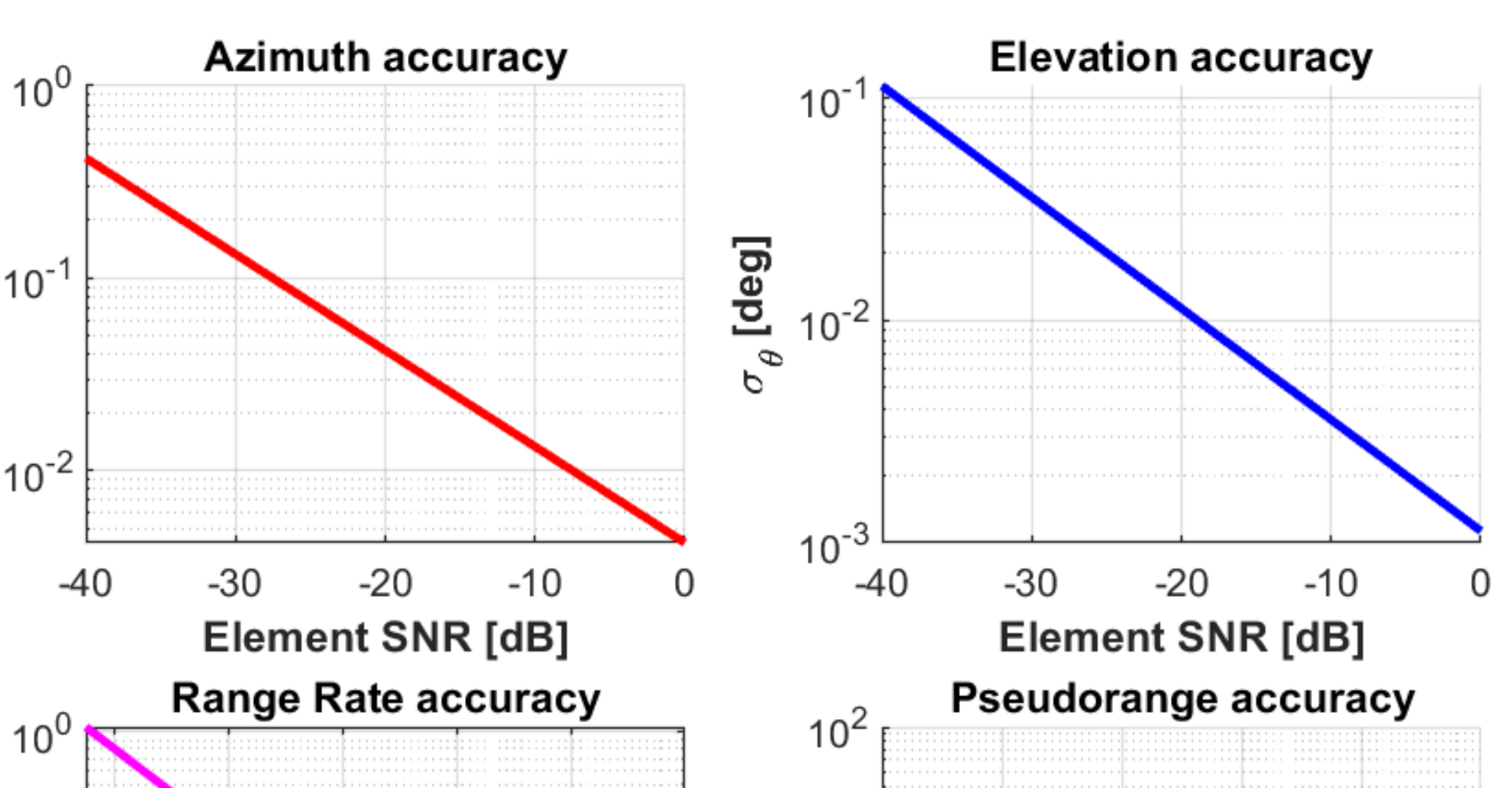

**Figure** 5. CRLB for a) Azimuth estimation (top-left), b) Elevation estimation (top-right), c) Range Rate estimation (bottom-left) and d) Pseudorange estimation (bottom-right)

### B. *Frequency Of Arrival*

Now let us focus on the accuracy of FoA estimation, which is performed using only the SSB. To minimize the CRLB for FoA, we consider a specific local replica waveform with a 10 ms duration, where the signal energy is concentrated at the beginning and the end of the window. In order words, FoA is estimated using two consecutive SSBs, as a single SSB does not have enough resolution for accurate estimation. This approach maximizes the effective root mean square duration (the time-domain equivalent of Gabor bandwidth for frequency estimation) compared to a waveform consisting only of a single PSS/SSS burst. This methodology is detailed in [18], where this local replica waveform is referred as *waveform* $k_2$.

The expected accuracy for the FoA estimation is assessed using the CRLB shown below:

$$\sigma_F = \frac{1}{2\pi} \cdot \frac{6}{\frac{C}{N_0} \cdot T^3 \cdot \left[\frac{\nu^2}{4} + 3\left(1 - \frac{\nu}{2}\right)^2\right]} \tag{3}$$

where $T$ is the coherent integration time and $\nu$ is the pulse duty cycle, i.e. the fraction of the 5G downlink frame containing the SSB, used for frequency estimation.

Furthermore, since real receivers can closely approach the CRLB as demonstrated in [18], we use the FoA CRLB as a realistic approximation of the expected accuracy for generating the FoA observables in this analysis. The subplot at the bottom left in Figure 5 illustrates the CRLB for FoA using only the SSB.

To maintain consistency with the navigation analysis, the FoA accuracy in Figure 5 is expressed in terms of range rate (m/s). From Figure 5, we can conclude that the range rate estimation error remains below 10 cm/s across the relevant $C/N_0$ range.

*C. Time Of Arrival*

Finally, we evaluate the accuracy of the ranging observable. In line with the frequency estimation approach, this evaluation is performed using only the SSB component of the downlink signal. To realistically assess Time of Arrival (ToA) estimation accuracy using the 5G SSB, we calculated the CRLB for a waveform consisting solely of the PSS and SSS. The analytical derivation of this CRLB is provided in Appendix B. This CRLB was derived by adapting the approach for pulsed signals proposed in [29] to the PSS+SSS combination, which has a signal bandwidth of approximately 15 MHz. As shown in the bottom-left subplot of Figure 5, the ToA accuracy computed for the PSS/SSS waveform in FR2 (with SCS = 120 kHz) remains below 30 cm across all relevant C/N0 regimes. Because real-world receiver performance can closely approach the CRLB under these conditions, as demonstrated in [18], it serves as a practical baseline for ToA accuracy in this study.

### 3.2 *Additional signal propagation impairments models*

*A. Elevation and Azimuth measurement impairments*

While the fundamental CRLB derived from thermal noise establishes an optimistic performance floor of approximately 0.1°-0.2° under nominal SNR conditions, a robust system analysis for a mobile Ka-band terminal must incorporate additional error sources. Due to the millimeter-scale wavelengths characteristic of the Ka-band, the system is exceptionally sensitive to beam misalignment, attitude uncertainty, heating-induced distortion, and tropospheric scintillation. Treating these independent error sources as orthogonal, their Root-Sum-Square (RSS) combination expands the realistic operational AoA tracking uncertainty to a standard deviation of about 0.3 [30]. Consequently, the pointing errors $\chi_\phi$ and $\chi_\theta$ are modeled as an independent normal distribution with a standard deviation of σ=0.3° on both azimuth and elevation.

*B. Pseudo-range and Doppler Impairments*

The primary error sources that must be accounted for in pseudorange and pseudorange-rate measurements are UE synchronization errors. The simulator models the local oscillator clock bias ($\Delta t$) and drift ($\Delta \dot{t}$) according to a target Temperature-Compensated Crystal Oscillator (TCXO) specification. The oscillator dynamics are represented using a random ramp model [28] where clock bias is randomly extracted and clock stability corresponds to a pseudorange rate variation of 10 m/s. Additive time/frequency noise characteristics representative of a standard TCXO are also included within the oscillator dynamic model [23].

The second impairment incorporated into the model is satellite ephemeris uncertainty. In 5G NTN, ephemerides are broadcast via the *ephemerisInfo* Information Element (IE) within SIB19 [31], with achievable accuracy depending on the ground infrastructure's Orbit Determination and Time Synchronization (ODTS) capability [23]. While Keplerian parameters offer millimetre-level spatial granularity suitable for extended high-accuracy ODTS, they require continuous numerical propagation at the UE. To minimize terminal computational burden during initial PRACH synchronization, the Position–Velocity (P–V) format is adopted, offering 1.3 m and 0.06 m/s spatial/velocity quantization step sizes compatible with meter-level accuracy provided by standard on-board GNSS receivers [32]. Following the Signal-in-Space Range Error (SISRE) framework [27], ephemeris error is modelled as a zero-mean Gaussian variable projected along the user line-of-sight, with an elevation-dependent standard deviation baselined at 1.5 m (position) and 0.1 m/s (velocity) at $5^{\circ}$ elevation [25]—aligning with the maximum P–V quantization step size [25].

The third modeled error source comprises satellite clock impairments, specifically synchronization and stability (Appendix A). Per 3GPP TS 38.104, Time Alignment Error (TAE) defines the maximum timing disparity across antenna connectors to preserve phase coherence and beamforming capability. The strict 65 ns limit for MIMO transmissions corresponds to a maximum pseudorange error of 19.5 m. Modeled as a $3\sigma$ bound, this translates to a Gaussian clock error with a standard deviation of $\sigma_{\Delta t_{Sat}} = 6.5\ m$. This error is elevation-independent, while clock stability is negligible and subsumed by the pseudorange-rate error terms ($\chi_{\dot{\rho}}$).

The fourth error source modelled in this work is the transmission chain bias, which accounts for the internal hardware group delays introduced by the satellite RF front-end payload electronics during beam forming and switching. As reported in [22], this impairment is modelled as a random constant drawn from a uniform distribution within [−30 ns, 30 ns] upon beam initialization and maintained strictly constant for the duration of the same Beam ID (BID). This range represents a worst-case scenario typical of an early-stage deployment, driven by unpredictable residual hardware phase offsets and group delays incurred during the payload electronics' switch-on and switch-off cycles of each beam without ground-provided calibration. Beyond its magnitude, investigating how these per-BID constant RF front-end biases propagate into the positioning solution throughout the satellite visibility window of a given beam, and evaluating techniques for their operational estimation or mitigation, represents a key research direction to enhance the positioning service.

The fifth error includes tropospheric error, and it was modeled according to [23]; this resulted in a residual error that varies from 10 cm at higher elevations to 80 cm according to elevation. For a given elevation, it is modelled as a Gaussian random variable. Ionosphere error in very high bandwidth like Ka can be assumed negligible. Additionally, multipath errors have been considered according to [34]. It is worth noting that the satellite tracking policy in Section 2-A mitigates multipath error by selecting signals with elevation angles above 40°, keeping multipath error below 2 m. Consequently, more complex modeling and analysis are unnecessary given the kilometer-level baseline error inherent to single-satellite constraints.

### 3.3 *Final Error Budget*

Table 3 summarizes the overall error budget defined for the simulation scenarios, while Figure 6 illustrates a representative realization of the resulting simulated performance. The right-hand panel depicts all generated observables over time, serving as the baseline input for the evaluations in Section 5. The left-hand panel details the corresponding individual impairments, demonstrating that the simulator incorporates the complete 5G-NTN error budget rather than relying on idealized receiver noise alone.

Table 3. Modelled additional error sources

| Contribution | Modelled error | Value | Source |
| --- | --- | --- | --- |
| | Azimuth and Elevation measurement | | |
| $\upsilon_{\phi}$ | Azimuth Estimation | 0.2° | [25] |
| $\upsilon_{\theta}$ | Elevation Estimation | 0.07° | [25] |
| $\chi_{\phi}$ | Azimuth Impairments | 0.3° | [25] |
| $\chi_{\theta}$ | Elevation Impairments | 0.3° | [25] |
| | *Pseudorange measurement* | | |
| $\upsilon_{\rho}$ | Ranging Thermal Noise | 0.3m | [18] |
| $\chi_{\rho}$ | Clock Bias | TCXO model | [23] |
| | Ephemeris error | 1.5 m @5°elevation | [23] |
| | Satellite clock error | 6.5 m | [27] |
| | Ionosphere error | negligible | [23] |
| | Troposphere error | 80 cm @5°elevation | [23] |
| | Transmission chain bias | $\mathcal{U}(-10m, 10m)$ | [22] |
| | Multipath | 2m @40°elevation | [34] |
| | *Pseudorange Rate Measurement* | | |
| $\upsilon_{\dot{\rho}}$ | Ranging Thermal Noise | 0.1m/s | |
| $\chi_{\dot{\rho}}$ | Clock Drift | TCXO model | [23] |
| | Ephemeris error | 0.1 m/s | [23] |
| | Satellite clock error | Negligible | [22] |
| | Ionosphere error<br>Troposphere error<br>Transmission chain bias<br>Multipath | 0.1 m/s | [23] |

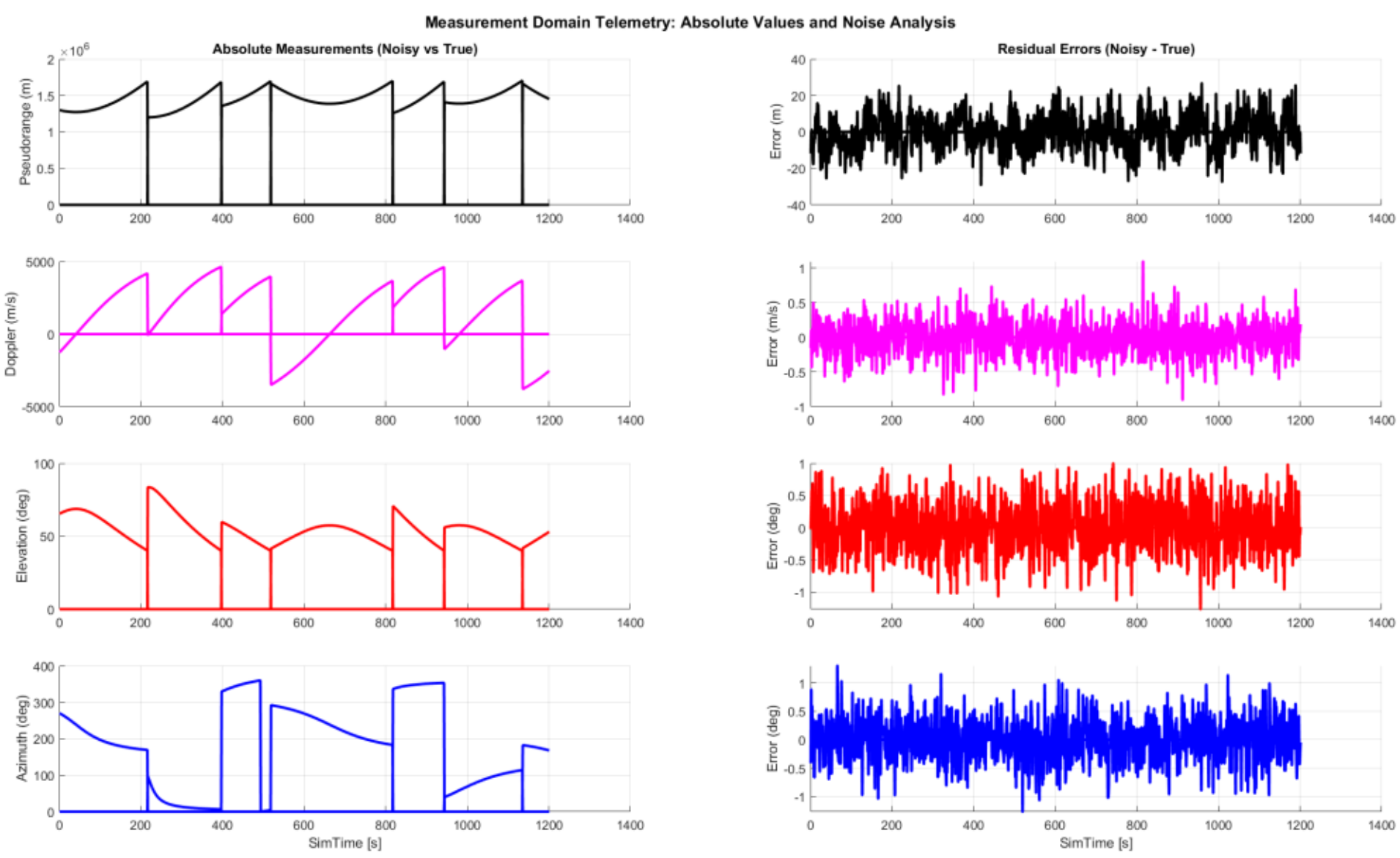

**Figure** 6. Example of observables generation with the application of related errors.

# 4 Multi-step Levenberg-Marquardt (LM) Positioning Algorithm

Even though the main objective of this paper is to enhance and streamline Random Access Channel (RACH) procedures for 5G NTN users, achieving the requisite positional awareness in GNSS-denied environments presents severe challenges. Standalone Angle-of-Arrival (AoA) tracking exhibits rapid spatial error growth proportional to altitude and angular noise, whereas Time-of-Arrival (ToA) and Frequency-of-Arrival (FoA) observations are heavily corrupted by unknown receiver clock bias and clock drift [11], [40]. To overcome these barriers without requiring multi-satellite visibility, several key additional contributions are introduced as enabling innovations. First, a multi-stage hybrid single-satellite localization pipeline is established: an initial spatial estimate derived purely from single-pass AoA bearing vectors yields a stable preliminary positioning estimate, which subsequently initializes expected propagation delays and frequency offsets so the state vector can be augmented with clock variables. Second, reconstructing and projecting 5G-NTN based pseudo-range and pseudo-range rate observation onto this single-satellite AoA baseline drastically tightens the spatial error ellipse, transforming the coarse single-satellite scenario into a tightly bounded solution [12], [41]. This multi-observable formulation is then improved by the inclusion of a Kalman Filter sequential estimation scheme allowing us to accumulate information over time and exploit geometry diversity. Standard Extended Kalman Filters frequently diverge during single satellite passes due to sparsity and ill-conditioned measurement Jacobians [42]. For this purpose, a Levenberg–Marquardt regularization within the Kalman update loop represent significant algorithmic advancements to eliminate numerical instabilities [43], [42] and it has been selected as the foundational machinery. The proposed LM-KF serves as an integrated framework that supports all stages of the multi-step positioning procedure, facilitating a seamless transition from an initial AoA-only coarse position estimate to a fully joint AoA/FoA/ToA multi-observable state solution. Furthermore, distinct filter configurations are implemented to enable the sensitivity analysis presented in the results section. The remainder of this section is organized as follows: first, a detailed mathematical formulation of the filter and its objective functions is presented, alongside an initial assessment of robustness against initialization errors; second, the integration of the filter within the multi-step workflow and its state-transition logic are described.

*4.1 Levenberg–Marquardt Kalman Filter implementation and setting*

The quadratic objective function to be minimized via Levenberg-Marquardt (LM) algorithm and integrate it within recursive sequential estimation is derived by minimization of generalized maximum a-posteriori (MAP) principle assuming Gaussian error distribution. The conventional approach to handling non-Gaussian errors, instead of adopting high-order methods like particle filters, is to over bound the unknown distribution with a boosted Gaussian model. While this renders the estimation suboptimal, it remains a widely accepted trade-off to minimize computational overhead. Under this assumption, the following equation applies:

$$\boldsymbol{J} = \left((\boldsymbol{x_t} - \boldsymbol{x_{t-1}})^T \boldsymbol{P}_{t,t-1}^{-1} (\boldsymbol{x_t} - \boldsymbol{x_{t-1}}) + \boldsymbol{r}(\boldsymbol{x})^T \boldsymbol{R}^{-1} \boldsymbol{r}(\boldsymbol{x})\right) \quad (4)$$

where $\mathbf{x}$ and $\mathbf{P}$ are the target state vector to be estimated and its covariance matrix, respectively. Instead, $\mathbf{P}_{t,t-1}$ is the a-priory state covariance matrix through the propagation step from state vector $\mathbf{x}_{t-1}$ at time $t-1$ to next epoch state vector $\mathbf{x}_t$, while $\mathbf{r}(\mathbf{x})$ is the residual vector, defined as the difference between the actual measurements $\mathbf{z}$ and the measurement model $\mathbf{h}(\mathbf{x})$:

$$\boldsymbol{r}(\boldsymbol{x}) = \boldsymbol{z} - \boldsymbol{h}(\boldsymbol{x}) \quad (5)$$

The term $\mathbf{R}$ represents the measurement covariance matrix associated with the current measurement set. Notably, the first term acts as a penalty incorporating information regarding the time correlation of the user positioning. As per [43], [42] the LM minimization is integrated within a Kalman filter framework [42] considering prediction-correction scheme of state vector and state covariance. In details:

1. *Measurement Update Step*

Considering availability of measurement $\boldsymbol{z}$ and a priori state estimation $\boldsymbol{x}_t^-$ to initialize the iterative process, the a-posteriori state update $\boldsymbol{x}_t^+$ at each iteration is found by solving the LM system:

$$\left(\boldsymbol{D}^T\boldsymbol{D} + \lambda diag(\boldsymbol{D}^T\boldsymbol{D})\right)\boldsymbol{\Delta x} = \boldsymbol{D}^T\boldsymbol{q}(\boldsymbol{x}) \quad (7)$$

where, $\boldsymbol{D}$ is the Jacobian matrix of $\boldsymbol{q}(\boldsymbol{x}) = \left\{\sqrt{\boldsymbol{R}^{-1}}\boldsymbol{r}(\boldsymbol{x}), \sqrt{\boldsymbol{P}_{t,t-1}^{-1}}\boldsymbol{p}(\boldsymbol{x})\right\}$ referred as extended residual vector. The square roots indicate Cholesky factorization of the fool measurement noise and state covariance matrices, $\mathbf{r}(\mathbf{x})$ is the residual model (5) and $p(\mathbf{x}) = (\mathbf{x}_t - \mathbf{x}_{t-1})$ is the state time update constraint derived from correspondent equation (6). λ is the damping parameter (Marquardt parameter). $\mathbf{\Delta x}$ is the step update given by $\mathbf{x_{k+1}} = \mathbf{x_k} + \mathbf{\Delta x}$, where $k$ indicates the iteration of LM algorithm. $\mathbf{x_k}$ **is** initialized with $\boldsymbol{x}_t^-$ and converge to $\boldsymbol{x}_t^+$. $\mathbf{P}_{t,t-1}$ can be also updated and through the iterations according to [42].

The covariance update step is instead managed by using standard Kalman Filtering update Joseph formula, so:

$$\boldsymbol{P}_{t|t} = (\boldsymbol{I} - \boldsymbol{K}_t\boldsymbol{H}_t)\boldsymbol{P}_{t|t-1}(\boldsymbol{I} - \boldsymbol{K}_t\boldsymbol{H}_t)^\top + \boldsymbol{K}_t\boldsymbol{R}_t\boldsymbol{K}_t^\top \quad (8)$$

Where the Kalman gain is

$$\boldsymbol{K_t} = \boldsymbol{P_{t|t-1}} \boldsymbol{H_t^\top} \left( \boldsymbol{H_t} \boldsymbol{P_{k|k-1}} \boldsymbol{H_*^\top} + \boldsymbol{R_t} \right)^{-1}$$

And measurement linear projector $\mathbf{H_t}$ is extracted from row of $\mathbf{D}$ correspondent to applicable measurement vector components.

2. Time Update Step

The state $\mathbf{x}_t$ is predicted in time from $\mathbf{x}_{t-1}$ according to standard Kalman filter time-update equations relevant to the specific use case, utilizing either a static or kinematic state transition matrix $\mathbf{F}$ [28], while accounting for the applicable process noise covariance matrix $\boldsymbol{Q}$.

$$\begin{cases} \boldsymbol{x_t} = \boldsymbol{f}(\boldsymbol{x_{t-1)}}) \\ \boldsymbol{P_{t,t-1}} = \boldsymbol{F} \boldsymbol{P_{t-1,t-1}} \boldsymbol{F}^T + \boldsymbol{Q} \end{cases} \tag{9}$$

Within this framework, it is also possible to regress to the standard LM algorithm by removing the time-update, reinitializing the state in Equation (9) at the next time step with an arbitrary first guess and removing the penalty in Equation (4), corresponding to $\boldsymbol{P_{t,t-1}^{-1}} = \boldsymbol{0}$. In the following this configuration will be referred with the expression "memoryless" LM solution.

The general LMKF described herein is specialized into four distinct estimator configurations, defined by the specific composition of the state vector, the selected measurement set, and the filtering parameters. This modular approach is designed to articulate the transition states within the execution framework described in the next section, while establishing the notation used in the simulation assessment to evaluate incremental performance gains across configurations. The four estimator configurations are defined as follows:

1. $J_0$, based on memoryless processing $of$ *AoA* measurements $\boldsymbol{z} = \{\tilde{\phi}, \tilde{\theta}\}$ and targeting positioning state estimate $\boldsymbol{x} = \{\lambda, \varphi\}|_{h=0}$.
2. $J_1$, based on filtered *AoA* measurements $\boldsymbol{z} = \{\tilde{\phi}, \tilde{\theta}\}$ targeting positioning state estimate of $\boldsymbol{x} = \{\lambda, \varphi\}|_{h=0}$.
3. $J_2$, based on filtered joint *AoA+FoA* measurements $\boldsymbol{z} = \{\tilde{\phi}, \tilde{\theta}, \tilde{\dot{\rho}}\}$ targeting positioning state estimate of $\boldsymbol{x} = \{\lambda, \varphi, \Delta\dot{t}\}|_{h=0}$.
4. $J_3$, based on filtered joint *AoA+FoA+ToA* measurements $\boldsymbol{z} = \{\tilde{\phi}, \tilde{\theta}\tilde{\dot{\rho}}, \tilde{\rho}\}$ and targeting positioning state estimate of $\boldsymbol{x} = \{\lambda, \varphi, \Delta t, \Delta\dot{t}\}\big|_{h=0}$.

Target state vectors include 2D user coordinates $\boldsymbol{x} = \{\lambda, \varphi\}|_{h=0}$, which refer to latitude and longitude respectively, and it can be extended to $\boldsymbol{x} = \{\lambda, \varphi, \Delta t, \Delta\dot{t}\}|_{h=0}$ in case of ranging and Doppler measurements augmentation, considering $\Delta t, \Delta\dot{t}$ respectively the user clock bias and drift. The $h = 0$ condition states that all the solution are constrained to Earth surface on datum for WGS84 reference frame since altitude is not directly observable and cannot be separated from clock biases. The solution can be extended to flying users by addition of external information about the altitude that can be supposed to be available for drones or aviation applications without affecting the generality of the results.

Considering incremental measurement set and state vector, the Jacobian $\boldsymbol{D}$ of $\mathbf{q}(\mathbf{x})$ can be always expressed as concatenation of the partial derivatives of each measurement type $\mathbf{h}(\mathbf{x})$ defined in Appendix A with respect to the target state vector $\boldsymbol{x}$. For the sake of completeness, further details regarding the core algorithmic implementation, including state space transformations, process and measurement noise covariance tuning, and filter covariance initialization strategies, are provided in Appendix C.

### *4.2 Multi-step procedure*

The details of such a multi-step procedure are provided in the following execution steps:

**Step 1: Initialization**

a) Sub-satellite point computation

The communication terminal receives the SIB19 broadcast from the traffic satellite and calculates the current sub-satellite point. This is possible considering broadcast information can be accessed by blind search of the communication terminal. It is worth mentioning that this method is only based on the exploitation of SSB, not requiring a VSAT in idle mode to process additional reference signals.

b) State initialization:

Utilize the calculated sub-satellite point longitude and latitude$\{\lambda_{Sat}, \varphi_{Sat}\}|_{h=0}$ as the initial estimate to feed the navigation solver.

c) Covariance initialization:

Initialize the covariance bounds to define the initial uncertainty of the state estimation from the expected admissible region around the sub-satellite point. This is furthermore discussed in Appendix-C.

**Step 2: Iterative optimization of AoA based LM algorithm**

d) Angle-of-Arrival based positioning estimation

Execute the Levenberg-Marquardt (LM) estimation algorithm proposed in 4.1 to iteratively refine the state estimate based on the AoA observables according to (4) and considering algorithm configuration $J_1$.

**Step 3: Conditional Doppler and pseudorange tracking**

e) In case FoA and ToA observations are enabled at receiver level by PSS/SSS sequences processing, $J_2$ Doppler and $J_3$ pseudorange based solution requires a further initialization step:

- Frequency-of-Arrival (FoA) initialization

We need to properly initialize the clock drift $\Delta\dot{t}_0$ . We use $J_1$ preliminary estimation $\{\lambda_{J1}, \varphi_{J1}\}\big|_{h=0}$ to extract a first guess of $\Delta\dot{t}_0$ by simply inverting the Doppler equations defined in Appendix-A:

$$\Delta\dot{t_0} = \frac{\dot{\tilde{\rho}} - \langle \boldsymbol{e}, \Delta\boldsymbol{v}\rangle\big|_{\{\lambda_{J1}, \varphi_{J1}\}=0}}{c} \tag{8}$$

According to Appendix-A, the error contribution we expect affecting FoA in (8) and the projection error of satellite ephemeris and coarse solution $\{\lambda_{J1}, \varphi_{J1}\}\big|_{h=0}$ are the ones that determine this coarse estimation of the drift. Therefore, from the error budget and covariance estimation of the solution $J_1$, a proper initialization of the initial covariance $P_{\Delta t_0}$ can also be provided.

- Time-of-Arrival (ToA) initialization

We use the AoA preliminary estimation $\{\lambda_{J1}, \varphi_{J1}\}\big|_{h=0}$ to reconstruct the pseudorange from the code-phase $\delta$ [35] solving code ambiguity[1] based on broadcast timing information and first guess $\{\lambda_{J1}, \varphi_{J1}\}\big|_{h=0}$. Then, we extract a first guess of $\Delta t_0$ by simply inverting the pseudorange equation defined in Appendix-A:

$$\Delta t_0 = \frac{\tilde{\rho} - \left\|\mathbf{X}_{Sat} - E2L(\lambda_{J1}, \varphi_{J1}, h = 0)\right\|}{c} \tag{9}$$

According to Appendix-A, the error contribution we expect affecting ToA and $\tilde{\rho}$ in (9) and the projection error of satellite ephemeris and coarse solution $\{\lambda_{J1}, \varphi_{J1}\}\big|_{h=0}$ are the ones that determine this coarse estimation of the bias. Therefore, from the error budget and covariance estimation of the solution $J_1$, a proper initialization of the initial covariance $P_{\Delta t_0}$ can also be provided.

**Step 4: Positioning Optimization handover**

f) As far as they are initialized, upon Doppler and Ranging observable activation, the solver can safely steer the $J_2$ and $J_3$ optimization, taking over $J_1$ at Step 2.

For the sake of validation, the algorithm has been tested across a wide range of different initialization points to assess its robustness against non-linearities and local minima. Figure 7 provides a visualization of an illustrative example to demonstrate the properties of the multi-epoch solution. Here, the target user is positioned at $60°, 0°$ and the algorithm is initialized at $0°, 60°$. In this scenario, it is important to highlight that memoryless $J_0$ estimator always restarts from the same initialization point without exploiting any temporal correlation. This approach exhibits certain instabilities and reveals ambiguities when the solution is initiated from unfeasible points. This test is intended solely to demonstrate the evolution of the algorithm across the search space; it is not applicable to the final approach presented hereafter, which leverages the ability to consistently improve algorithm initialization within the neighborhood of the solution by utilizing knowledge of the current approximated sub-satellite point. This strategy drastically reduces the search space, ensuring that $J_0$ also converges to the proper minimum without exhibiting multiple solutions.

[1] Note that for a LEO satellite at a 1200 km altitude altitude, the signal propagation delay remains strictly below the 5G downlink frame duration of 10 ms. This observation is valid for user equipment (UE) maintaining an elevation angle greater than 10° relative to the satellite. Considering this, the integer code ambiguity is always 0, so that we can set up the pseudorange $\tilde{\rho}$ measurement from the measured code phase $\delta$. Without loss of generality, it is assumed that for higher altitude satellites, the accuracy of $\{\lambda_{J1}, \varphi_{J1}\}\big|_{h=0}$ and the currently considered error on ephemeris are considered sufficient to discriminate the unknown code ambiguity and build $\tilde{\rho}$.

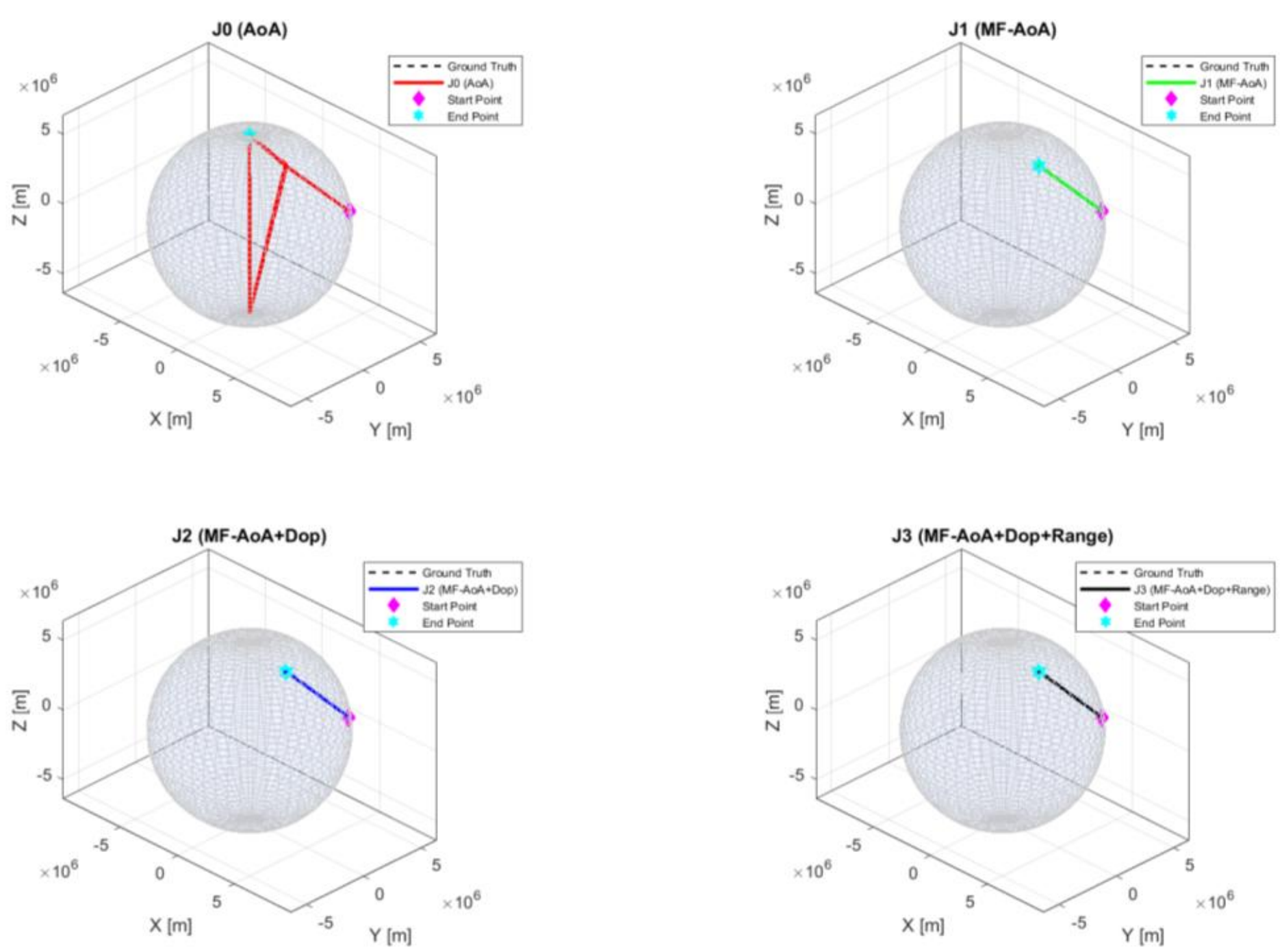

**Figure** 7. Convergence of different algorithms over the Earth surface.

## 5 Simulation results

This section presents numerical results from our simulation-based analysis across two main use cases: (i) static user scenarios evaluated over a global geographic grid spanning representative latitudes and longitudes, and (ii) a dynamic scenario based on a realistic vehicular trajectory around Milan, Italy. Furthermore, overall performance is evaluated in terms of location uncertainty and its direct impact on successfully completing the RACH procedure.

### 5.1 *Static user simulations*

The numerical simulation results for static evaluation tests have been conducted over a predefined global grid. The geographic grid covers latitudes from $0°$ to $70°$ at $10°$ increments, and four discrete longitudes spanning $0°$ to $180°$ with a $60°$ spacing. Simulations are executed over a $1200$ s time window, during which the full satellite constellation trajectory is propagated. The serving satellite is dynamically assigned throughout this interval according to local target site visibility and the handover strategy outlined in Section 2. Navigation observables are sampled at $1$ Hz without loss of generality, adhering strictly to the measurement models and system/processing impairments specified in Section 3. These generated observables are subsequently evaluated by the integrated positioning engine across the four distinct estimator configurations ($J_0$, $J_1$, $J_2$, and $J_3$). This multi-stage evaluation enables a systematic assessment of the incremental performance gains achieved when transitioning from memoryless snapshot processing to multi-epoch, multi-observable filtering architectures. Finally, a sensitivity analysis is carried out by scaling the angle-of-arrival (azimuth and elevation) measurement impairments $\chi_{\phi,\theta}$ and noise $\upsilon_{\phi,\theta}$ conditions by factors of 1/3 (Best Scenario), 1 (Nominal Scenario), and 3 (Worst Scenario).

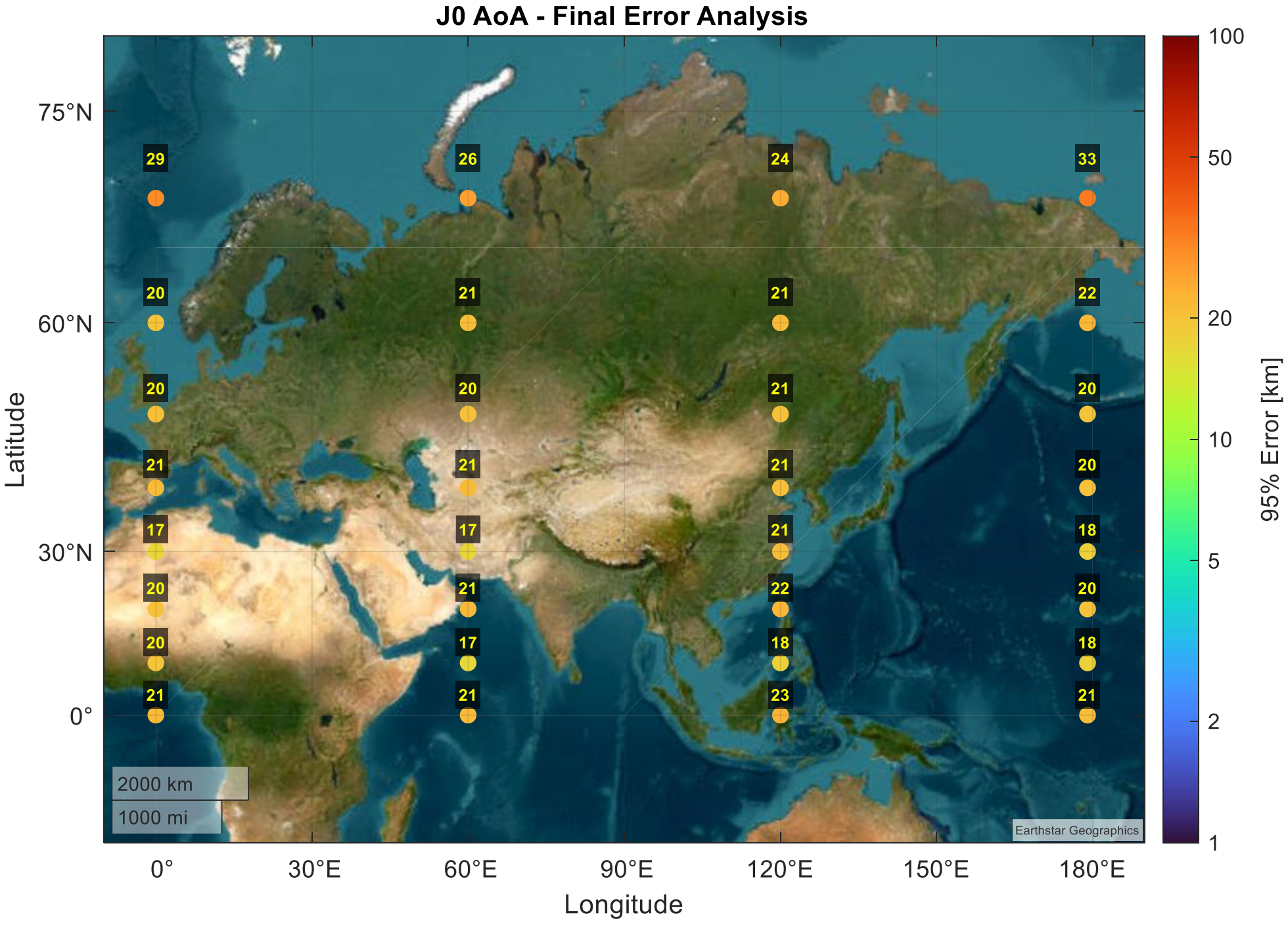

**Figure 8.** Estimation accuracy for J0 fixed locations grid for Nominal Scenario

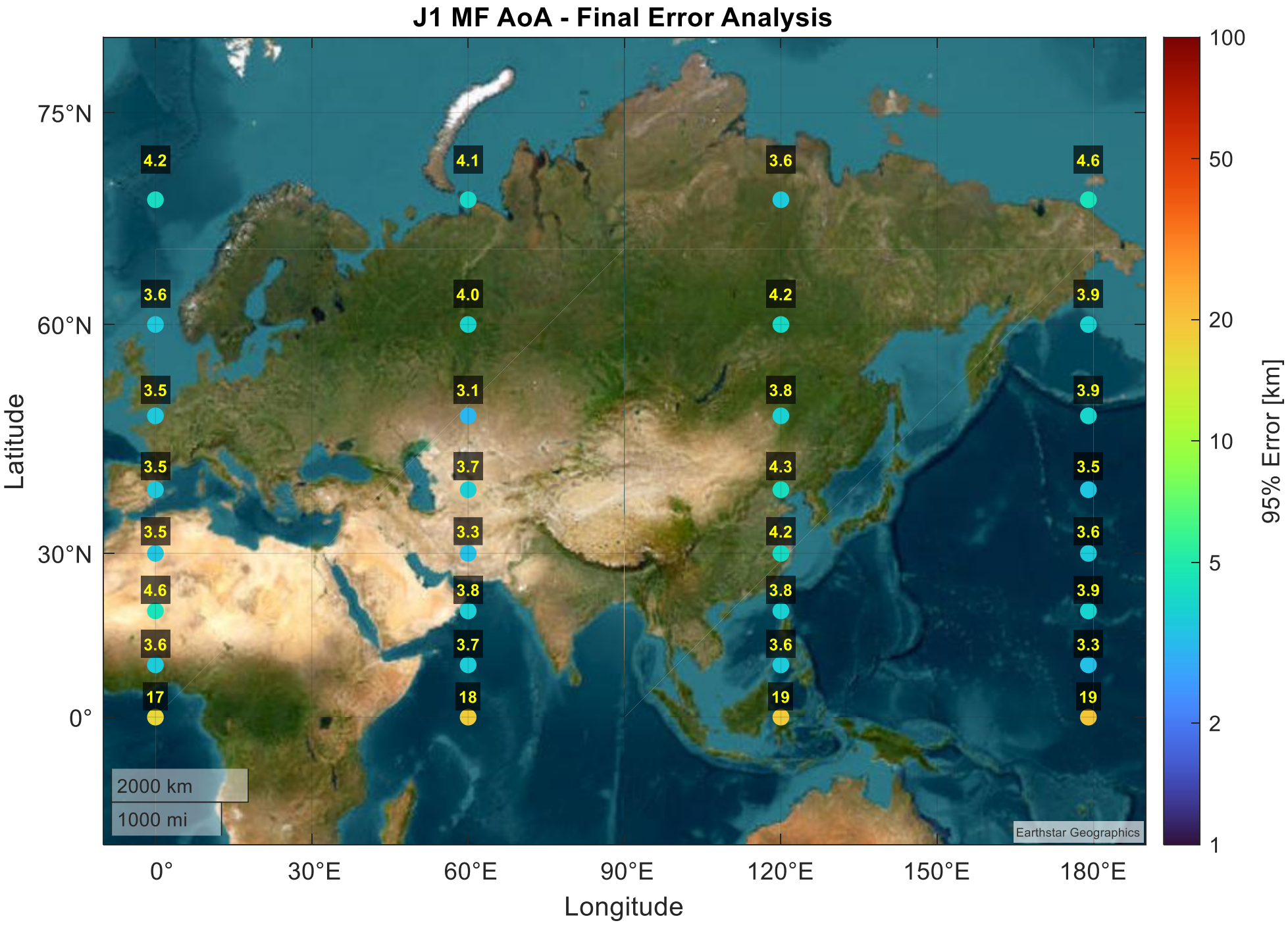

**Figure 9.** Estimation accuracy for J1 fixed locations grid for Nominal Scenario

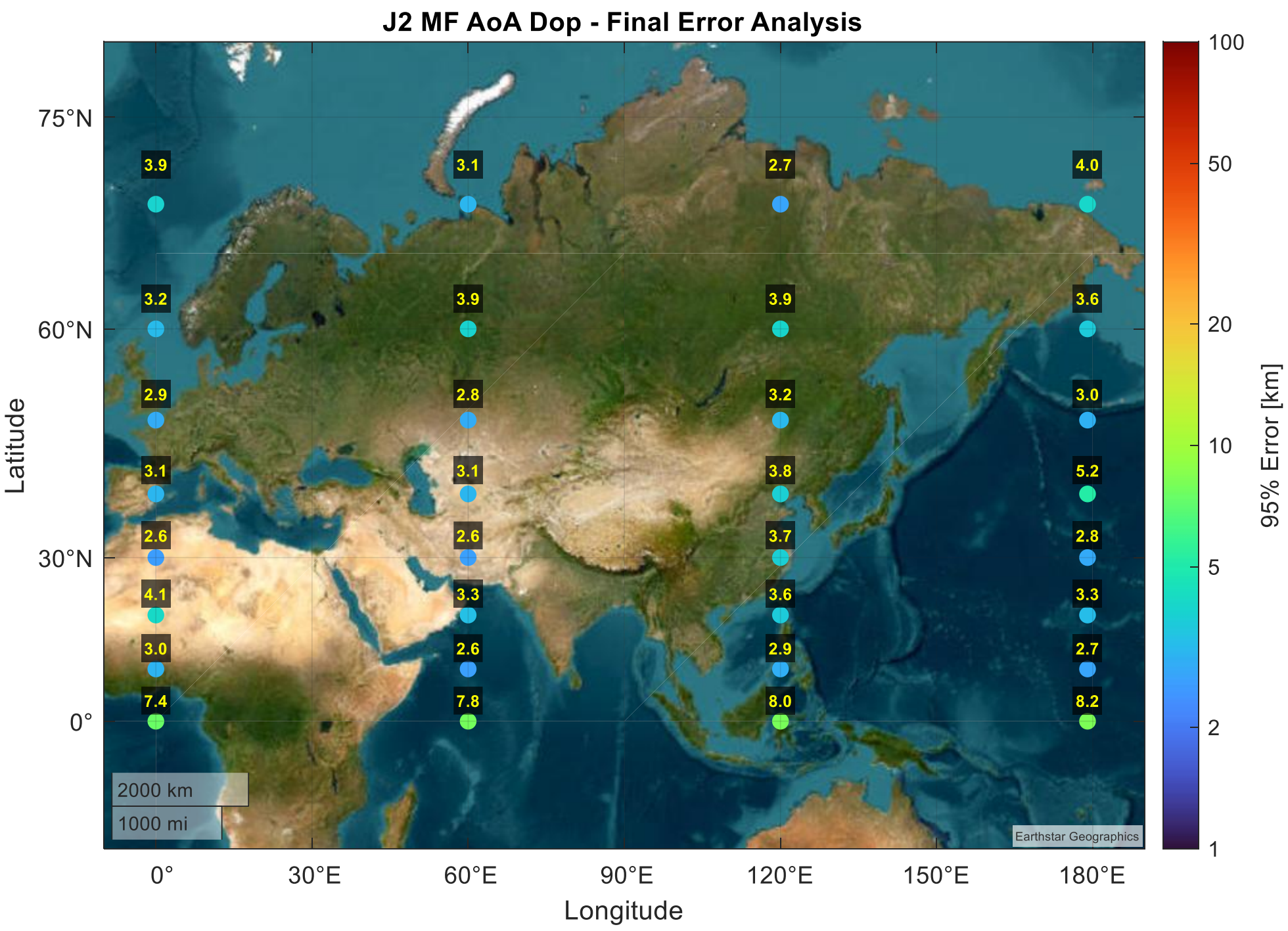

**Figure 10.** Estimation accuracy for J2 fixed locations grid for Nominal Scenario

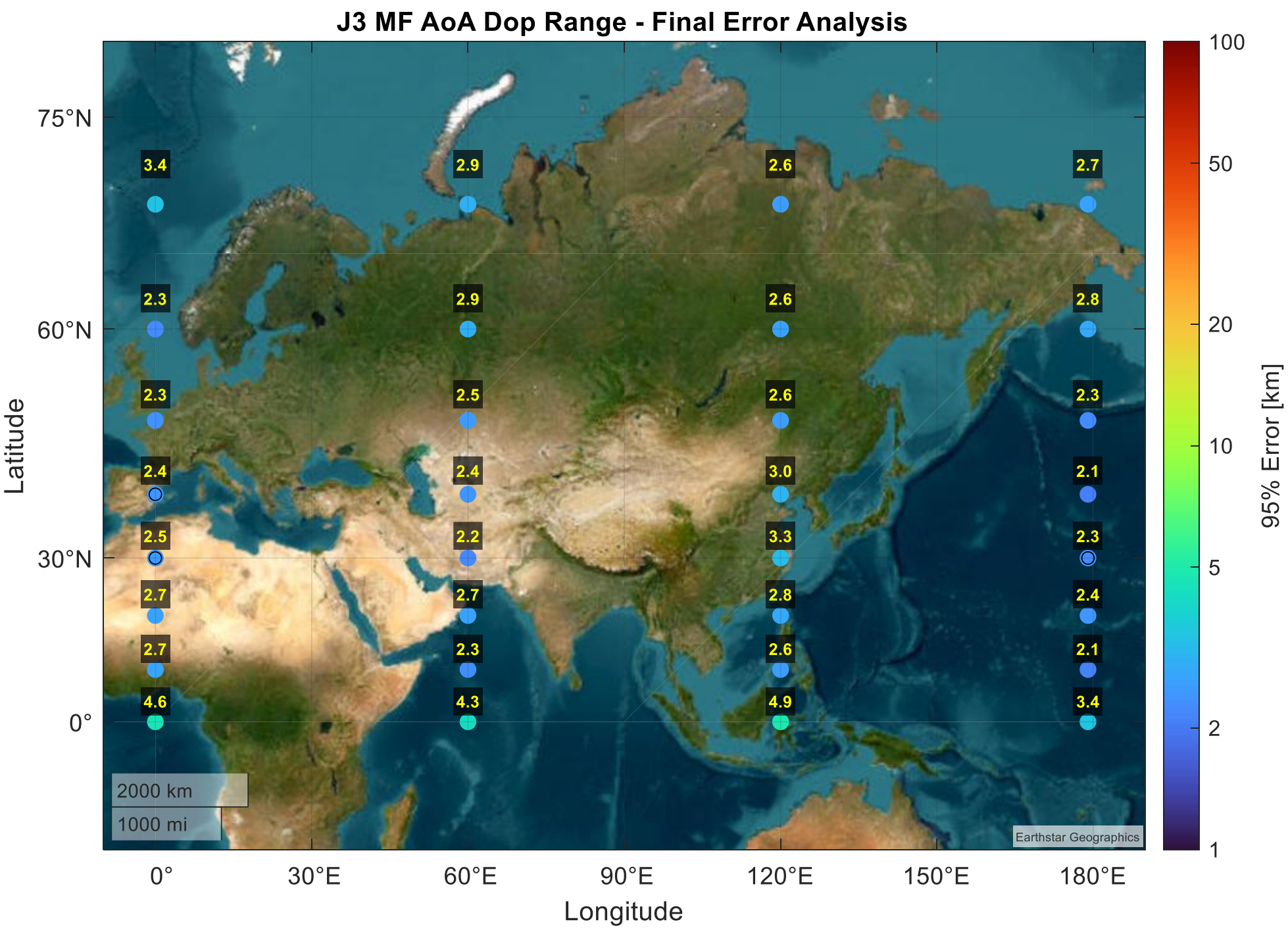

**Figure 11.** Estimation accuracy for J3 fixed locations grid for Nominal Scenario

These results are visually illustrated in Figures 10 through 13 for the nominal scenario. Notably, the $J_0$ estimator consistently produces errors exceeding 15 km across all evaluated grid points. Operating without temporal correlation while relying solely on single-epoch AoA observables renders the solution highly susceptible to adverse instantaneous geometry, leading to pronounced positional inaccuracies. Conversely, the $J_1$ filter yields significantly lower errors by leveraging multi-epoch geometry diversity, effectively filtering out most of the measurement error. Specifically, nominal errors remain below 5 km across most grid points, except near the equator (0° latitude), where $J_1$ filtering offers limited improvement. At low latitudes, ground receivers experience reduced satellite pass diversity: the inclined satellites considered in the target 5G NTN constellation traverse low-latitude regions at flatter, lower-elevation geometries compared to mid-latitudes, while polar satellites exhibit wider longitudinal spacing at the equator. Consequently, when the highest visible satellite near the equator exhibits a relatively low elevation angle, relying exclusively on AoA tracking yields poor multi-epoch geometric dilution of precision in the multi-epoch estimate.

To further enhance accuracy, integrating pseudorange rate and pseudorange observables is essential to introduce supplementary geometric diversity. The inclusion of Doppler data in the $J_2$ configuration allows the filter to stably constrain errors below 10 km, achieving performance with less than 3 km error at optimal locations. Nevertheless, the primary performance breakthrough stems from the pseudorange constraints introduced in $J_3$. By incorporating pseudorange measurements, the $J_3$ estimator overcomes the geometric vulnerabilities inherent to purely angular and Doppler tracking, consistently driving global positioning errors below 5 km.

Table 4 summarizes global statistical metrics across the geographic grid for all three measurement quality scenarios. For completeness, full spatial visualizations over the grid are provided in Appendix D.

**Table 4. Statistics measured over the grid point errors for each solver**

| Worst Scenario | | | | |
|---|---|---|---|---|
| **Metric** | $J_0$ | $J_1$ | $J_2$ | $J_3$ |
| Mean error (km) | 62.12 | 16.82 | 11.41 | 8.90 |
| Median error (km) | 62.15 | 11.37 | 9.59 | 8.23 |
| Std. Deviation (km) | 7.36 | 14.58 | 4.83 | 2.17 |
| Max Error (km) | 87.66 | 57.49 | 24.42 | 14.69 |
| Min Error (km) | 50.65 | 9.28 | 7.80 | 6.25 |

| Nominal Scenario | | | | |
|---|---|---|---|---|
| **Metric** | $J_0$ | $J_1$ | $J_2$ | $J_3$ |
| Mean error (km) | 21.23 | 5.92 | 4.01 | 2.83 |
| Median error (km) | 20.54 | 3.78 | 3.31 | 2.62 |
| Std. Deviation (km) | 3.64 | 4.90 | 1.75 | 0.67 |
| Max Error (km) | 32.73 | 19.42 | 8.19 | 4.94 |
| Min Error (km) | 16.82 | 3.09 | 2.58 | 2.11 |

| Best Scenario | | | | |
|---|---|---|---|---|
| **Metric** | $J_0$ | $J_1$ | $J_2$ | $J_3$ |
| Mean error (km) | 7.02 | 1.87 | 1.38 | 1.10 |
| Median error (km) | 6.85 | 1.27 | 1.10 | 0.98 |
| Std. Deviation (km) | 1.09 | 1.62 | 0.68 | 0.41 |
| Max Error (km) | 11.00 | 6.50 | 3.07 | 2.48 |

| Min Error (km) | 5.60 | 1.03 | 0.88 | 0.70 |
|---|---|---|---|---|

The sensitivity to AoA accuracy is immediately apparent: degrading the angular error by a factor of three (Worst Scenario) prevents any estimator configuration from reaching sub-5 km accuracy. Nevertheless, multi-observable augmentation plays an even more crucial role in constraining total error bounds across all configurations. Conversely, under the Best Scenario, enhancing AoA precision serves as a primary driver to achieve 1 km-level positioning performance. Although the relative improvements from Doppler and pseudorange constraints are less transformative when angular error is low, these additional observables remain fundamental in bounding maximum peak errors within this regime.

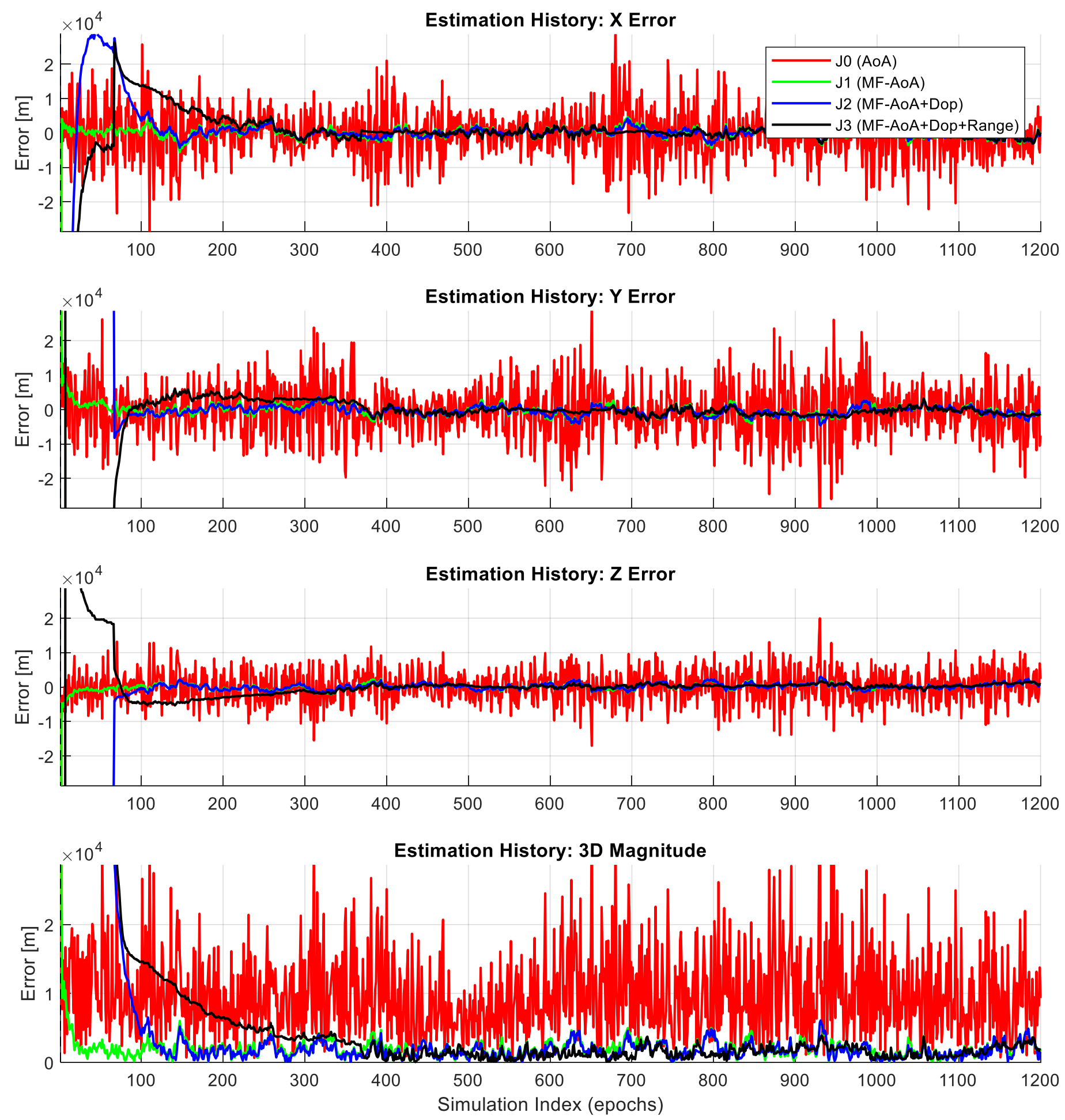

**Figure 12. Steady state error for each implemented solver for the grid point with coordinates** $(60^{\circ}, 60^{\circ})$

To evaluate the temporal convergence of the multi-step positioning strategy, a time-series analysis at a representative geographic location provides valuable insights. Figure 12 illustrates the time-domain evolution of the LMKF across the four estimator configurations ($J_0$ through $J_3$). A distinct transient phase is observable during the transition from $J_1$ to the augmented $J_2$ and $J_3$ solutions. This demonstrates the effectiveness of the sequential estimation procedure in properly initializing the clock bias and drift states, a prerequisite for leveraging the geometric projections of Doppler and pseudorange observables. While detailed transient optimization falls outside our current scope, these results highlight a key trade-off: filtering large initial errors via multi-epoch temporal smoothing incurs a convergence delay that depends heavily on local trajectory geometry and satellite handover dynamics. Consequently, optimizing these two factors will be a primary driver for real-world implementation

### *5.2 Kinematic user simulation*

The dynamic simulation results have been evaluated over a realistic land-vehicle trajectory derived from a Joint Research Centre (JRC) measurement campaign [44]. The primary objective is to assess positioning performance under dynamic receiver conditions while relying on a single serving satellite signal. The evaluation tests the capability of each LMKF setting to track a ground vehicle traveling at speeds up to 100km/h over a 50min observation window. For the sake of brevity this scenario has been tested considering Worst Scenario setting for the measurement generation. Figure 13 illustrates the downsampled estimated position scatterplots in local North-East coordinates across all four positioning estimations, superimposed on the ground truth vehicle trajectory (black line) with designated start (green circle) and end (red star) markers. The estimated position points are color-mapped chronologically according to time epoch 0-3000s to highlight trajectory tracking capability over time. As observed, spatial dispersion around the ground truth narrows substantially when moving from solver $J_0$ to $J_3$. Specifically, the snapshot $J_0$ AoA solution exhibits severe point scattering over the 25 km without temporal coherence. Multi-epoch filtering in $J_1$ and $J_2$ progressively tightens scatter around the true motion profile, aligning the color gradient with vehicle propagation. Ultimately, the $J_3$ solver achieves the tightest clustering along the true trajectory, demonstrating superior 2D horizontal tracking accuracy under dynamic conditions.

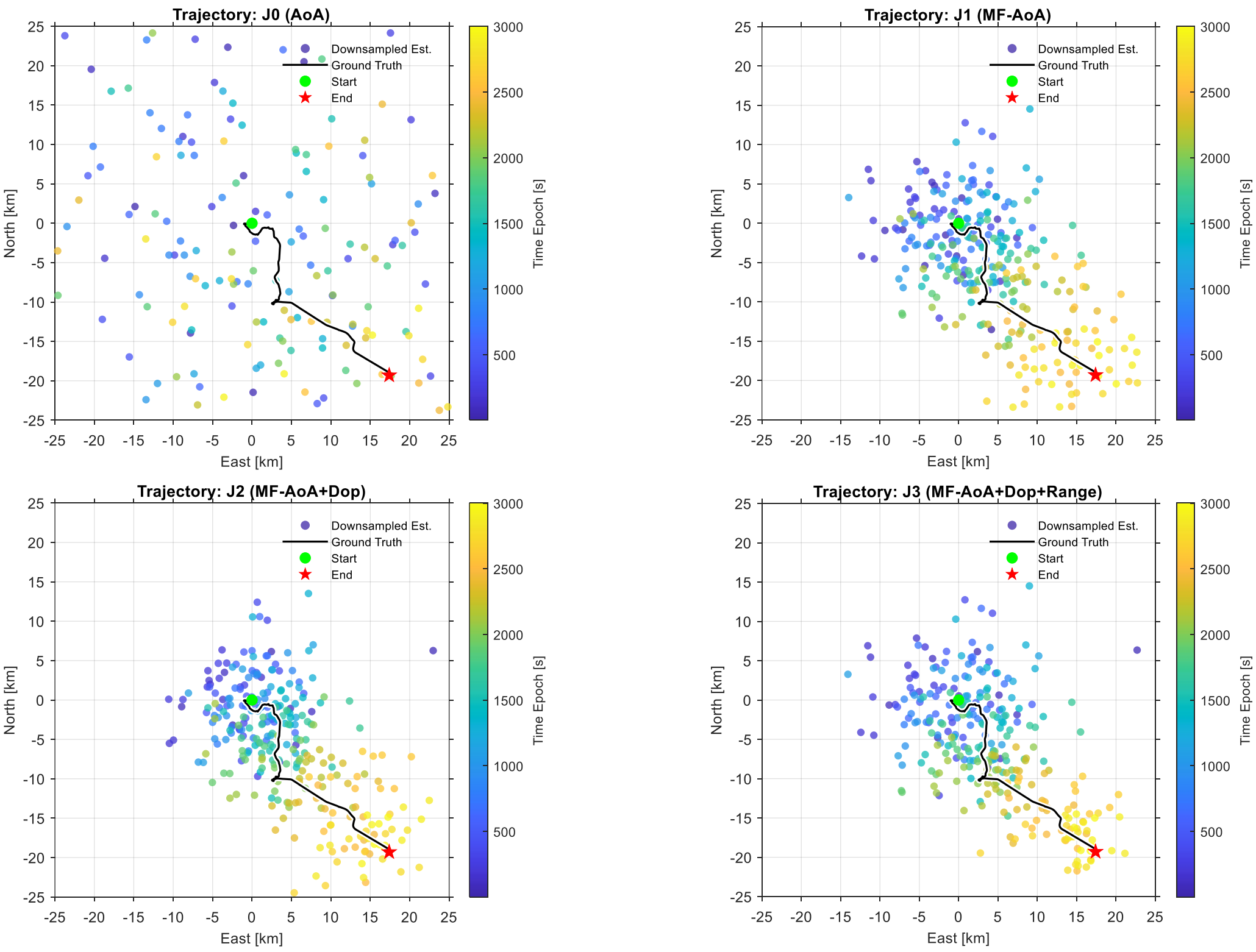

**Figure 13.** 2D error for each algorithm in the dynamic trajectory: (top-left) 2D error for $J_0$, (top-right) 2D error for $J_1$, (bottom left) 2D error for $J_2$ and (bottom right) 2D error for $J_3$.

**Table 5.** Statistics measured over the dynamic simulation errors for each solver

| Metric | $J_0$ [km] | $J_1$ [km] | $J_2$ [km] | $J_3$ [km] |
|---|---|---|---|---|
| 2D error (95th perc) | 61.88 | 11.71 | 10.30 | 9.50 |

Table 5 presents the 95th percentile errors along the trajectory for all four solvers. Moving from $J_0$ to $J_3$achieves a statistically significant error reduction, consistent with Figure 13. Crucially, the error levels align with expected baseline performance, demonstrating that performance holds when transitioning from static to slowly moving users.

### 5.3 *Location uncertainty discussion*

This subsection focuses on the possibility of successfully completing RACH procedure in GNSS denied environment, when the UE location uncertainty is reduced using the single-satellite-based solver, presented in this paper.

The residual location uncertainty translates into a differential delay and Doppler. Specifically, as reported in [14], differential delay is the difference between the one-way delay from the true UE position to the satellite and the one-way delay from the erroneous location to the satellite. Various location uncertainties, corresponding to different maximum horizontal errors, were investigated in [14]. Due to geometric properties, horizontal position errors do not translate linearly into differential delay; specifically, position errors yield significantly lower differential delays at higher elevation angles compared to lower elevation angles (e.g., 30°).

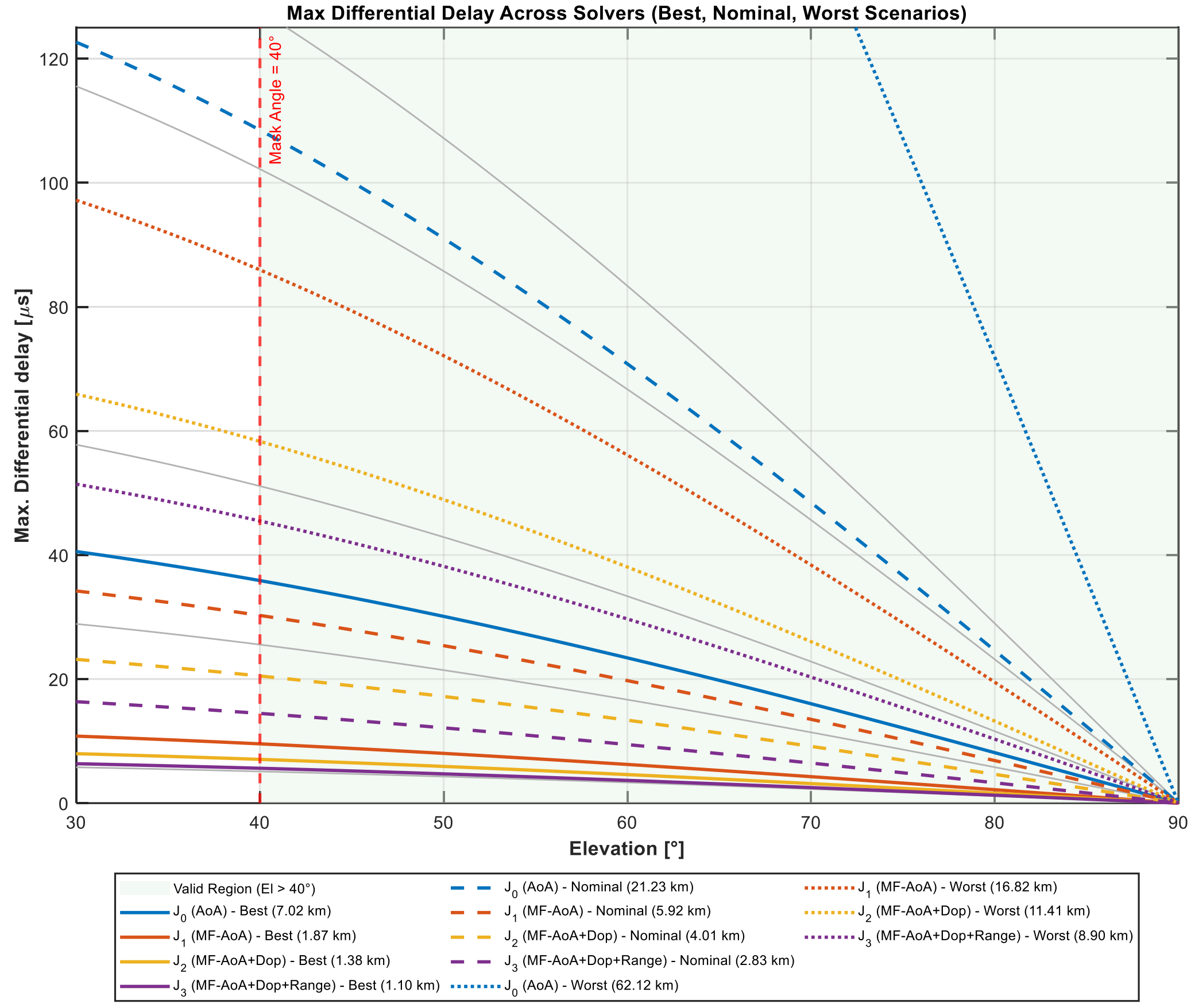

**Figure 14.** Maximum differential delay vs elevation angles for different values of location uncertainty radii.

Regarding differential Doppler, [14] shows that a 25 km location uncertainty induces a maximum differential Doppler of 1.038 ppm, translating to a 31.14 kHz shift at 30 GHz. Coherently integrating over a 31.14 kHz shift during the symbol duration (8.92 $\mu$s at numerology = 3) does not introduce any shift in the Zadoff-Chu correlation peak, confirming that differential Doppler has a negligible impact on PRACH detection.

Figure 14 provides a comprehensive evaluation of how horizontal position uncertainties coming from Table 4 map into maximum differential delays, corresponding directly to timing advance errors, as a function of the satellite elevation angle (30° to 90°). In accordance with our satellite selection and handover policy, the operational domain is established for elevation angles above 40°, highlighted as the valid region in Figure 14. The curves explicitly illustrate the performance progression across our multi-step solver hierarchy ($J_0$ through $J_3$) under Best, Nominal, and Worst operational scenarios. At the 40° policy boundary, the baseline Angle-of-Arrival solver ($J_0$) exhibits high residual delay, exceeding $100\,\mu$s in Nominal and Worst cases. Incorporating matched-filter measurements ($J_1$) tightens this uncertainty significantly, reducing the Nominal differential delay to $\sim 30\,\mu$s at 40° elevation and dropping steadily at higher angles. Integrating Doppler ($J_2$) and Range ($J_3$) observations achieves sub-3 km position accuracy in Nominal conditions (2.83 km for $J_3$) and down to 1.10 km in the Best scenario. Within the valid $> 40°$ elevation domain, the differential delay for $J_2$ and $J_3$ remains exceptionally low, dropping below $20\,\mu$s even under Worst scenario bounds and continuously decreasing as geometry improves.

Currently, 3GPP specifications define stringent timing advance tolerances for PRACH preamble reception at the network side as well as UE transmission requirements. For FR2-NTN ranges with an SCS of 120 kHz, the differential roundtrip delay limits range from $0.65\,\mu$s (PRACH format B3) to $8.11\,\mu$s (PRACH format C2). These stringent constraints practically preclude completing the PRACH procedure in GNSS denied environment.

For this reason, RAN4 is reviewing these requirements, distinguishing between UEs operating with GNSS and those under GNSS-denied conditions. The location uncertainties currently under investigation in the RAN4 Work Item description (WID) span from 5 km to 45 km.

The solution proposed in this paper aligns with this direction, by reducing the location uncertainty from the full beam radius down to few kilometres, thereby requiring a minimal relaxation of the differential delay requirements.

As shown in Figure 14, while these strict restrictions preclude standard PRACH procedures under high baseline positioning uncertainties, operating within the bounds of our satellite selection policy combined with our localization framework opens vital operational avenues in GNSS-denied environments. Specifically, the $J_3$ configuration allows the UE to successfully complete the PRACH procedure under current 3GPP requirements for elevation angles greater than 60° in the Best configuration (i.e., highest AoA accuracy). Similarly, $J_2$ configuration can successfully complete the PRACH procedure at elevation angles greater than 70°.

However, a minor relaxation of requirements is necessary to ensure PRACH finalization at any elevation angle above 40° for both the Nominal and Best configurations.

Future work will build upon this foundation, exploring dynamic orchestration frameworks that exploit our localized position estimates to optimize seamless mobility and connection continuity across dense LEO constellations.

## 6 Conclusions

This paper introduced a novel, integrated single-satellite multi-step positioning framework designed to reduce localization uncertainty and mitigate the PRACH initialization limitations present in current 5G NTN specifications for high-frequency (Ku/Ka-band) VSAT terminals. Operating in GNSS-denied environments, these terminals cannot currently leverage multi-satellite reception; however, they possess the key advantage of using phased-array antennas to derive AoA measurements. While traditional AoA-only techniques are constrained to coarse accuracies on the order of dozens of kilometers, the proposed approach initializes position estimation via single receive-beam AoA steering, which is then used to reliably reconstruct ToA and FoA observables derived from standard 5G synchronization signals (PSS/SSS). These multi-observable measurements are processed by a custom LMKF estimation engine. By exploiting user time-correlation and sequential satellite geometry changes, the multi-step framework bridges the gap between coarse directional tracking and fine spatial resolution, delivering significantly improved positioning accuracy.

The performance of the framework has been validated through extensive static global grid simulations incorporating high-fidelity measurement models for AoA, FoA, and ToA. Sensitivity analyses conducted across different cost-function configurations highlight the incremental contributions of each observable, transitioning from single-epoch angular tracking to multi-epoch filtering, followed by the integration of Doppler and ranging observables. Under nominal error conditions, unfiltered AoA processing yields large, highly dispersed position estimation errors that consistently exceed a maximum error of 30 km. Applying LMKF filtering to AoA tracking successfully leverages motion physics to smooth random measurement noise, bringing the global mean positioning error below 20 km. However, the true game-changing performance, eliminating massive error spikes at equatorial latitudes, comes from the introduction of FoA and ToA observations. Introducing Doppler shift tracking serves as a major milestone by incorporating relative velocity dynamics, which drops the mean error below the 10 km threshold. Ultimately, full multi-observable hybridization with ToA ranging bounds the maximum position error below 5 km. The analysis has also been extended from static to slowly dynamic users by considering realistic 50-minute vehicle trajectory profiles, demonstrating that the framework achieves a similarly tight, highly reliable steady-state position error along dynamic trajectory paths.

This paper proposes a solution that reduces location uncertainty from the entire beam radius to a few kilometers, requiring only a minimal relaxation of differential delay requirements. While current specifications preclude standard PRACH procedures under high baseline positioning uncertainties, combining the proposed satellite selection policy with our localization framework enables successful operation without GNSS across several evaluated configurations. Other configurations, e.g. those with lower AoA accuracy, require a minor relaxation of requirements to ensure successful PRACH completion.

Future work will extend this single-satellite framework to address non-Line-of-Sight urban blockages and optimize multi-epoch filter initialization across broader sub-satellite boundaries. Additionally, blind AoA acquisition and tracking algorithms will be evaluated under analogue or hybrid beamforming architectures, which offer the most practical path for UE implementation. More realistic satellite selection policies, including fixed-time handover scheduling, will also be explored to improve location uncertainty reduction and refine navigation solution transients. Finally, future efforts will bridge the gap between theoretical models and practical implementation by assessing the computational efficiency of LKMF.

## Appendix A

This appendix details mathematical models for the different observables generated within the system and user terminal simulator and then referenced in the positioning algorithm.

*I. AoA – Azimuth and Elevation model*

$$\begin{cases} \phi = atan2(e_{los}, n_{los}) \\ \theta = atan2\left(\sqrt{e_{los} + n_{los}}, u_{los}\right) \end{cases} \quad \text{(A.1)}$$

The azimuth observable $\phi(\text{x})$ and elevation $\theta(\text{x})$ are derived from the line-of-sight unit vector components $\{e_{los}, n_{los}, u_{los}\}$ expressed in ENU reference frame (see Appendix D) representing the relative position vector between ECEF position of the user $\boldsymbol{X_{user}}$ and ECEF position of the satellite $\mathbf{X_{Sat}}$ retrieved by SIB message demodulation. Indicating with $z = \{\tilde{\phi}, \tilde{\theta}\}$ the correspondent measurements, the full measurement model can be expressed as :

$$\begin{cases} \tilde{\phi} = \phi + \chi_{\phi} + v_{\phi} \\ \tilde{\theta} = \theta + \chi_{\theta} + v_{\theta} \end{cases} \quad \text{(A.2)}$$

Where: $\chi_{\phi}$ $\chi_{\theta}$ are respectively the systematic imperilments affecting azimuth and elevation measurements and $v_{\phi}$ and $v_{\theta}$ are gaussian white noise representing residual receiver processing errors derived in Section 3 and simulated in Section 5.

*II. Pseudo-range and Pseudo-range rate models*

$$\begin{cases} \dot{\rho} = \langle \boldsymbol{e}, \Delta \boldsymbol{V} \rangle + c(\Delta t - \Delta t_{Sat}) \\ \rho = \|\Delta \boldsymbol{X}\| + c(\Delta \dot{t} - \Delta \dot{t}_{Sat}) \end{cases} \quad \text{(A.3)}$$

The $\rho(\text{x})$ and $\dot{\rho}(\text{x})$ are respectively referred as pseudo-range and pseudo-range rate observable [27] derived from receiver signal processing and reconstruction procedure from code-phase and Doppler raw measurements defined in Section 4. The $\boldsymbol{e} = \Delta \boldsymbol{X}/\|\Delta \boldsymbol{X}\|$ is the LOS director cosines expressed in ECEF. The geometrical ranging and ranging rate observables are primarily affected by synchronization and clock stability errors at UE and satellite side. $\Delta t$ and $\Delta \dot{t}$ represent respectively the clock bias and drift due to the UE local oscillator; $\Delta t_{Sat}$ and $\Delta t_{Sat}$ are the expected constellation synchronization. Their generation has been discussed in Section 3.

In (A.3) $\Delta \boldsymbol{X}$ and $\Delta \boldsymbol{V}$ are respectively expressed as

$$\begin{cases} \Delta \boldsymbol{X} = \mathbf{X}_{\boldsymbol{Sat}} - \boldsymbol{X_{user}} \\ \Delta \boldsymbol{V} = \boldsymbol{V_{Sat}} - \boldsymbol{V_{user}} \end{cases} \quad \text{(A.4)}$$

Where $\boldsymbol{V_{user}}$ is the velocity vector associated to the target user. At this stage we can similarly define the target measurement equation as

$$\begin{cases} \tilde{\dot{\rho}} = \dot{\rho} + \chi_{\dot{\rho}} + \upsilon_{\dot{\rho}} \\ \tilde{\rho} = \rho + \chi_{\rho} + \upsilon_{\rho} \end{cases} \tag{A.5}$$

where $\chi_{\dot{\rho}}$ and $\chi_{\rho}$ are respectively the systematic components of pseudo-range rate and pseudo-range errors and $\upsilon_{\dot{\rho}}$ and $\upsilon_{\rho}$ are gaussian white noise error components derived in Section 3 and simulated in Section 5.

## Appendix B

This appendix aims at computing the CRLB for the TOA estimation for an OFDM-based signal with continuous frequency allocation around DC subcarrier. This case applies to 5G PSS/SSS. For an OFDM signal, the received signal on the $k$-th subcarrier can be written as

$$r_k = \sqrt{C} s_k e^{-j2\pi k \Delta f \tau} + n_k \tag{B.1}$$

where $\tau$ is the unknown delay to estimate, $s_k$ is the transmitted modulation symbol on the $k$-th subcarrier, $C$ is the transmit power, $\Delta f$ is the SCS for the OFDM symbol and $n_k$ is the thermal noise modelled as zero-mean, white Gaussian stochastic complex process with variance $\sigma_N^2$. It is worth noting that $\sigma_N^2$ is the variance of the single noise process component, i.e. the noise power $P_N = 2\sigma_N^2$. Let us recall that $k$ spans from $-N$ to $N$, where for the 5G PSS/SSS case $N = 63$, so that the total number of allocated subcarriers is 127.

The first derivative of the mean value of $r_k$ is given by

$$\dot{m}_k(\tau) = \frac{\partial m_k(\tau)}{\partial \tau} = -j2\pi k \Delta f \sqrt{C} s_k e^{-j2\pi k \Delta f \tau} \tag{B.2}$$

The analytical expression of the unsmoothed CRLB for an OFDM-based signal with continuous frequency allocation can be derived from the Fisher information matrix given by:

$$I(\tau) = \left[ \frac{1}{\sigma_N^2} \dot{\boldsymbol{m}}^H(\tau) \cdot \dot{\boldsymbol{m}}(\tau) \right] \tag{B.3}$$

where $\dot{\boldsymbol{m}}(\tau)$ is the first derivative of the mean value of the received signal for all the active subcarriers and $(.)^H$ indicates the Hermitian operator. By including equation (B.2) into equation (A.3), we obtain:

$$I(\tau) = \frac{4\pi^2 \Delta f^2 C}{\sigma_N^2} \sum_{k=-N}^{N} k^2 \cdot |s_k|^2 \tag{B.4}$$

Assuming that $|s_k|^2$ is unitary, and recalling that

$$\sum_{k=1}^{N} k^2 = \frac{N \cdot (N+1) \cdot (2N+1)}{6} \tag{B.5}$$

Equation (A.4) can be rewritten as

$$I(\tau) = \frac{4\pi^2 \Delta f^2 C}{\sigma_N^2} \cdot 2 \cdot \frac{N \cdot (N+1) \cdot (2N+1)}{6} \tag{B.6}$$

After simple manipulations of (A.6), we obtain:

$$I(\tau) = \frac{8\pi \Delta f^2 \cdot SNR \cdot N \cdot (N+1) \cdot (2N+1)}{3} \tag{B.7}$$

The CRLB in meters for a single OFDM symbol with a continuous subcarrier allocation centered at DC is given by:

$$\sigma_\tau \geq \sqrt{\frac{3 \cdot c_0}{8\pi \Delta f^2 \cdot SNR \cdot N \cdot (N+1) \cdot (2N+1)}} \tag{B.8}$$

where $c_0$ is the speed of light. This result can be considered as the open-loop performance or the unsmoothed one, following the terminology from [33]. The variance of the smoothed TOA estimate at the output of a code-tracking loop properly designed can be written as

$$\sigma_{sm}^2 \approx \sigma_\tau^2 \cdot 2B_L T \tag{B.9}$$

where $B_L$ is the loop bandwidth, whereas $T$ is the loop update time, corresponding in the specific case of 5G PSS to one frame, i.e. 10 ms.

It is worth noting that the thermal noise regime considered here lies in the region where the receiver performance achieves the theoretical bounds, justifying the use of the CRLB rather than the Ziv-Zakai bound (ZZB). Furthermore, for ToA, reference [18] benchmarked receiver performance and theoretical bound, demonstrating that DLL tracking nearly reaches the CRLB limit.

## Appendix C

*I. Coordinate Transformation and Maps*

Considering position vector all models assume the following state variable conversion:

A) Latitude, Longitude, Height to Earth Centered Earth Fixed Transformation LLH2ECEF, indicated as E2L which includes according to [27] or [35] full transformation matrix and linear mapping

$$\boldsymbol{X} = \{X, Y, Z\} = E2L(\varphi, \lambda, h) = E2L(\mathbf{x})$$

$$\boldsymbol{T_{L2E}} = \frac{\partial \boldsymbol{X}}{\partial \boldsymbol{x}} \tag{C.1}$$

B) Earth Centered Earth Fixed to East, North Down reference frame Transformation ECEF2ENU, indicated as E2ENU which includes according to [27] or [35] full transformation matrix and linear mapping

$$\boldsymbol{e} = \{e, n, u\} = \text{E2ENU}\ (X, Y, Z, \varphi_0, \lambda_0, h_0) = \text{E2ENU}(\mathbf{X}, \mathbf{x_0})$$

$$\boldsymbol{T_{L2E}} = \frac{\partial \boldsymbol{e}}{\partial \mathbf{X}} \tag{C.2}$$

In C.2 $X, Y, Z$ is the target position in ECEF and $\varphi_0, \lambda_0, h_0$ is the point where reference frame is referenced to.

C) Latitude, Longitude, Height to East, North Down LLH2ENU indicated as L2ENU which includes according to [27] or [35] full transformation matrix and linear mapping

$$\boldsymbol{e} = \{e, n, u\} = L2\text{ENU}\ (\varphi, \lambda, h, \varphi_0, \lambda_0, h_0) = \text{E2ENU}(\mathbf{x}, \mathbf{x_0})$$

$$\boldsymbol{T_{L2E}} = \frac{\partial L2\text{ENU}}{\partial \boldsymbol{x}} \tag{C.3}$$

In (C.2) $\varphi, \lambda, h$ is the target position in LLH and $\varphi_0, \lambda_0, h_0$ is the point where reference frame is referenced to.

The proposed framework explicitly accounts for the inverse transformation and the corresponding inverse linear mapping during processing $\{e_{los}, n_{los}, u_{los}\} = \text{E2ENU}(\mathbf{X_{Sat}}, \mathbf{x})$.

*II. Measurement equation handling and Jacobian details*

Appendix A equations apply to the complete model $\mathbf{h}(\mathbf{x}) = \{\phi, \theta, \dot{\rho}, \rho\}$**.** It is worth noting that the residual vector generally represented as per (5) has been better expressed for the elevation and azimuth observables according to the model (C.5).

$$\mathbf{r}_{\phi,\theta} = atan2\left(sin(\tilde{\theta} - \theta), cos(\tilde{\phi} - \phi)\right) \tag{C.4}$$

Considering (7), the Jacobian of the model can be better expressed as block matrix:

$$\mathbf{D} = \begin{bmatrix} \sqrt{\mathbf{R}^{-1}}\mathbf{H} \\ \mathbf{P}_{t,t-1}^{-1} \end{bmatrix} \tag{C.5}$$

Where $\mathbf{H} = \frac{\partial \mathbf{h}(\mathbf{x})}{\partial \mathbf{x}}$ . Considering as target state vector $\boldsymbol{x} = \{\lambda, \varphi, \Delta t, \Delta \dot{t}\}|_{h=0}$ and specializing for each configuration the observation model $\mathbf{h}(\mathbf{x})$ follows**:**

A. From AoA measurement model complete derivative to compute the correspondent sub-block of matrix $\mathbf{H}$ in equation (C.6) have been derived with respect to state components:

$$H_{(\phi,\theta)} = \begin{bmatrix} \frac{\partial \phi}{\partial \mathbf{e}}\frac{\partial \mathbf{e}}{\partial \mathbf{X}}\frac{\partial \mathbf{X}}{\partial \varphi} & \frac{\partial \phi}{\partial \mathbf{e}}\frac{\partial \mathbf{e}}{\partial \mathbf{X}}\frac{\partial \mathbf{X}}{\partial \lambda} & 0 & 0 \\ \frac{\partial \theta}{\partial \mathbf{e}}\frac{\partial \mathbf{e}}{\partial \mathbf{X}}\frac{\partial \mathbf{X}}{\partial \varphi} & \frac{\partial \theta}{\partial \mathbf{e}}\frac{\partial \mathbf{e}}{\partial \mathbf{X}}\frac{\partial \mathbf{X}}{\partial \lambda} & 0 & 0 \end{bmatrix} \tag{C.6}$$

B. From FoA and ToA model complete derivative to compute the correspondent sub-block of matrix $\mathbf{H}$ in equation (C.7) can be derived with respect to state components :

$$\mathbf{H}_{\dot{\rho}\rho} = \begin{bmatrix} \frac{\partial \dot{\rho}}{\partial \mathbf{X}}\frac{\partial \mathbf{X}}{\partial \varphi} & \frac{\partial \dot{\rho}}{\partial \mathbf{X}}\frac{\partial \mathbf{X}}{\partial \lambda} & 0 & 1 \\ \frac{\partial \rho}{\partial \mathbf{X}}\frac{\partial \mathbf{X}}{\partial \varphi} & \frac{\partial \rho}{\partial \mathbf{X}}\frac{\partial \mathbf{X}}{\partial \lambda} & 1 & 0 \end{bmatrix} \quad \text{(C.7)}$$

Where $\partial \mathbf{e}$ and $\partial \mathbf{X}$ derivatives are extracted from coordinate transformation maps defined above.

It is worthy to remark that we did not consider velocity of user in the target state vector and its derivatives considering that for the targeted implementation the assumption of quasi-stationarity of the user applies The extension to drones and aviation can be contemplated within this algorithm as far as the information about the receiver speed is provided by an external sensor (i.e. low-grade inertial sensor). Nevertheless, as demonstrated in the section 5 the model is completely applicable for a wide range of dynamic users whose dynamic is negligible with respect to LEO $\|\boldsymbol{V}_x\| \ll \|\boldsymbol{V}_{Sat}\|$.

*III. State Transition Matrix and process/measurement covariance setting*

As concern the state transition model of position state $\{\varphi_t\ \lambda_t, h_t\} = \mathbf{f}(\varphi_{t-1}\ \lambda_{t-1}, h_{t-1})$ the approximation of static or quasi-static user is handled by considering pure random walk process $\mathbf{f} = \mathbf{I}_{[3x3]}$ where the magnitude of process covariance $\mathbf{Q}^{llh}$ is used to properly regulated time correlation in the filtering architecture with $\mathbf{Q}^{llh} = \boldsymbol{T}_{ENU2L}\mathbf{Q}^{ENU}\boldsymbol{T}_{L2ENU}^{T}$ with $\mathbf{Q}^{ENU}(\mathbf{3},\mathbf{3}) = \mathbf{0}$. In case of clock bias and drift augmentation the state transition is completed with clock model which is the standard random ramp model in [28], [36].

In the end is worthy to remind that setting of measurement covariance matrix $\mathbf{R}$ has been considered as explained in Section 4.1 by proper Gaussian overabounding of the combination of estimated measurement noise and systematic impairments.

*IV. Setting of initial covariance*

For the sake of completeness, regarding the initialization of the LMKF algorithm in terms of the initial covariance matrix $\mathbf{P}_0$, the methodology described in Section 4.2 is directly inherited. At Step 1(c), an admissible region is defined around the selected initial sub-satellite point. As illustrated in Figure 2, the sub-satellite point defines the initial state $\mathbf{x}_0 = [\lambda_0, \varphi_0]^T$ and a region bounding $[-20°, 20°]$ in both latitude and longitude, according to a conservative minimum user elevation angle of $20°$. This bound is subsequently used to define a circular variance for the target state vector components $\mathbf{x}_0$. Moving to the full state covariance $\mathbf{P}_0$—including clock bias and clock drift for configurations $J_2$ and $J_3$ —then requires implementing the initialization procedure proposed in Step 3(e).

## Appendix D

This appendix presents the numerical values of the position errors for static scenario in case of Worst and Best Scenarios.

.

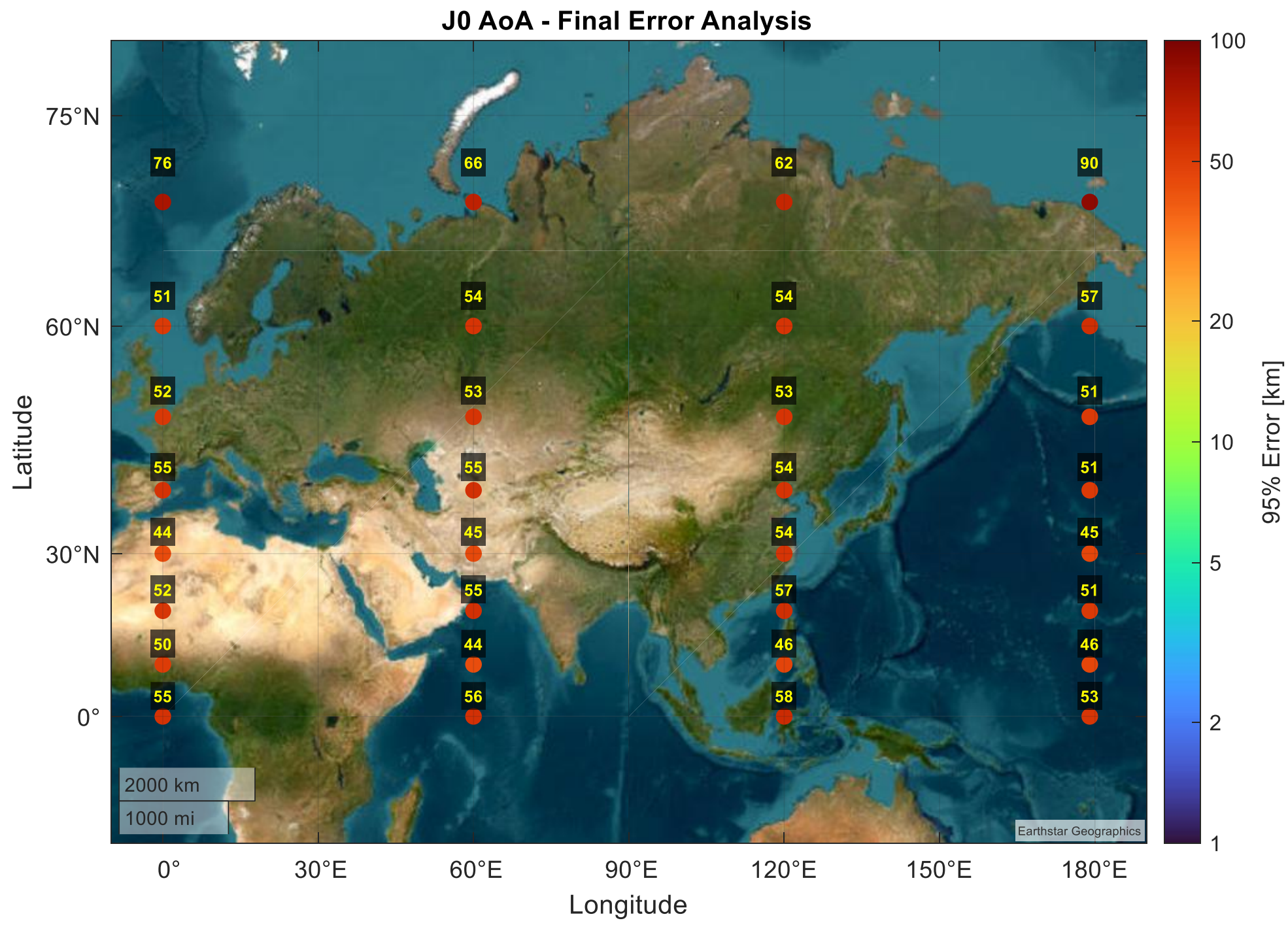

**Figure 15.** Estimation accuracy for J0 fixed locations grid for Worst Scenario

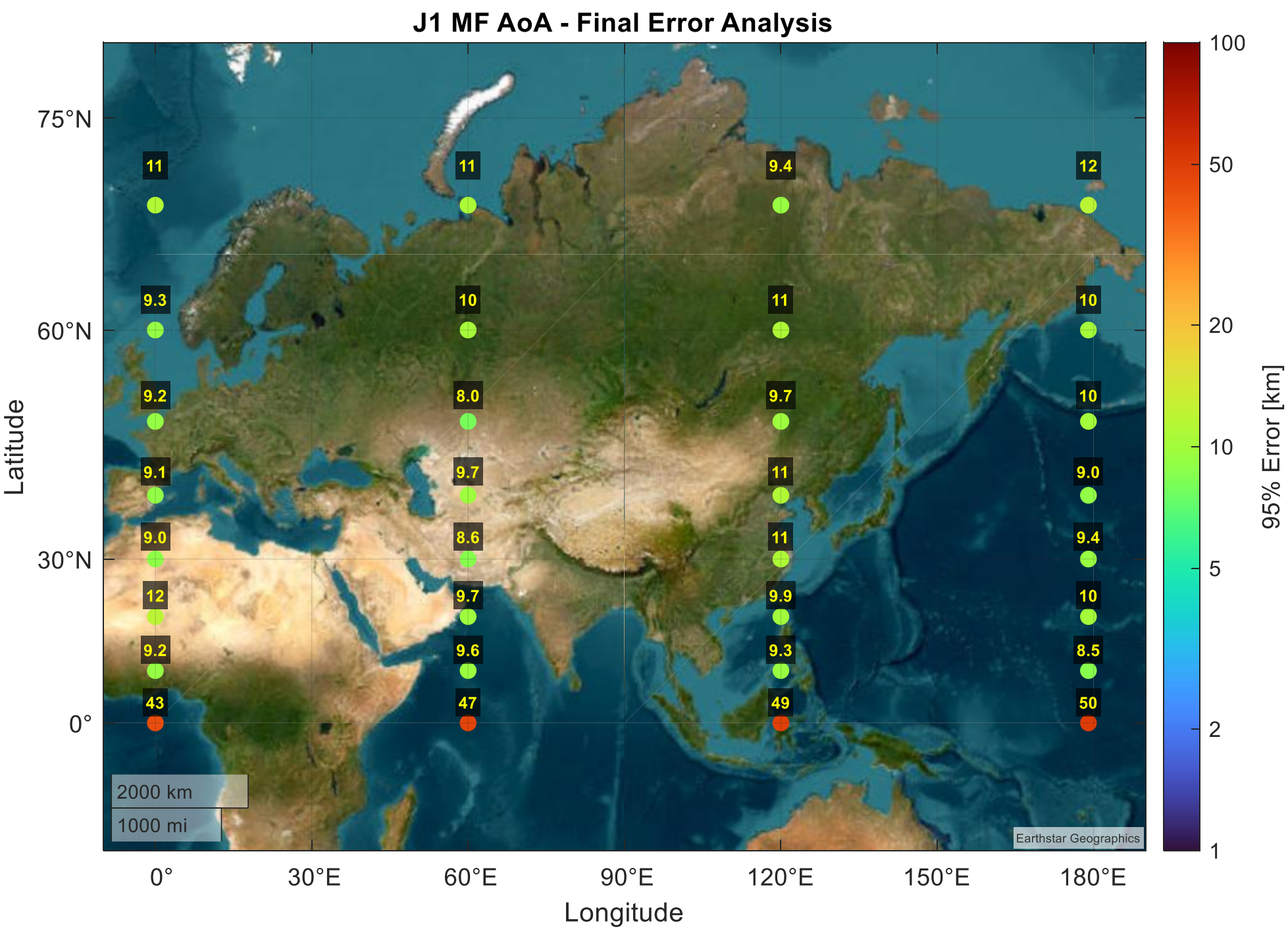

**Figure 16.** Estimation accuracy for J1 fixed locations grid for Worst Scenario

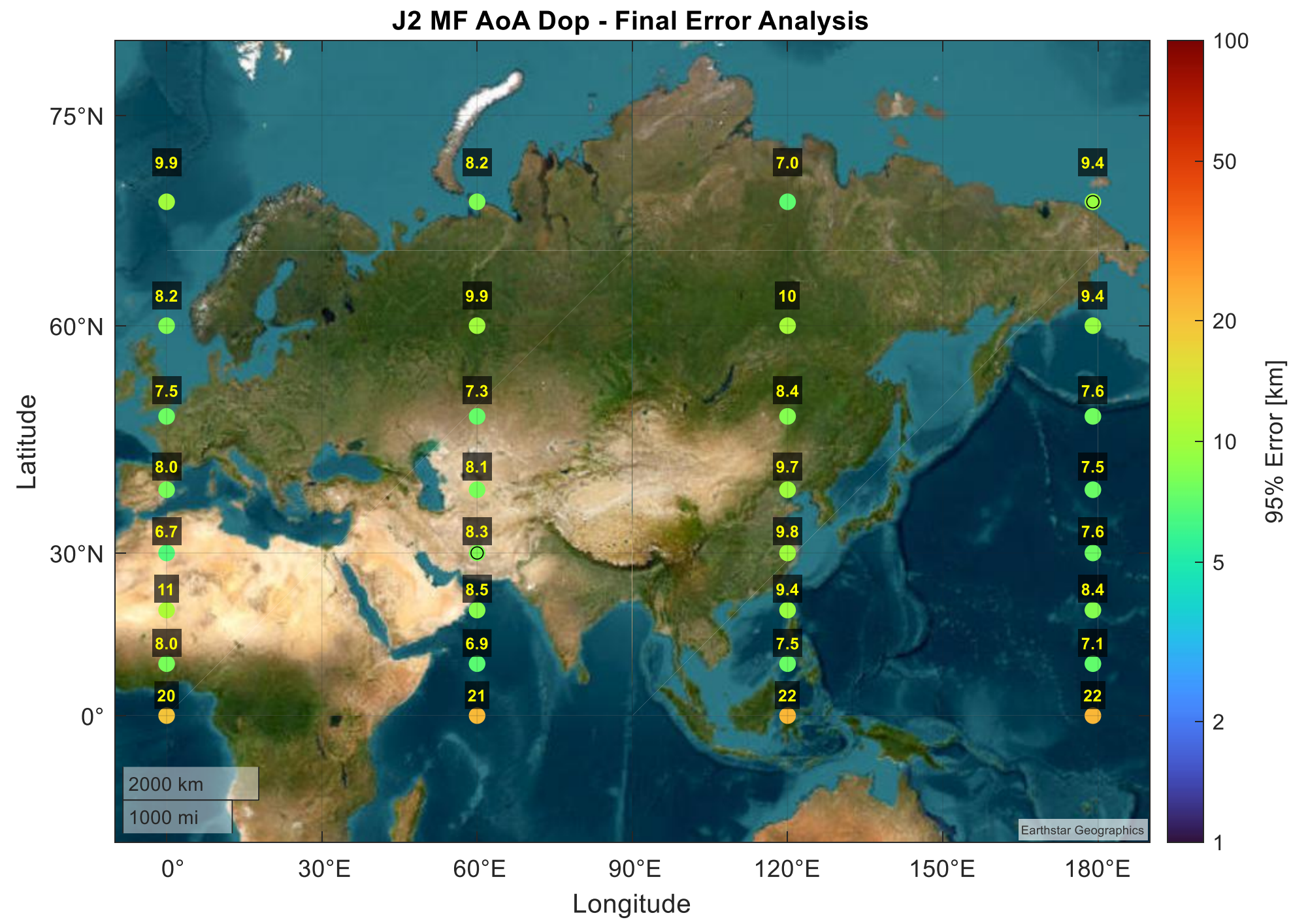

**Figure 17.** Estimation accuracy for J2 fixed locations grid for Worst Scenario

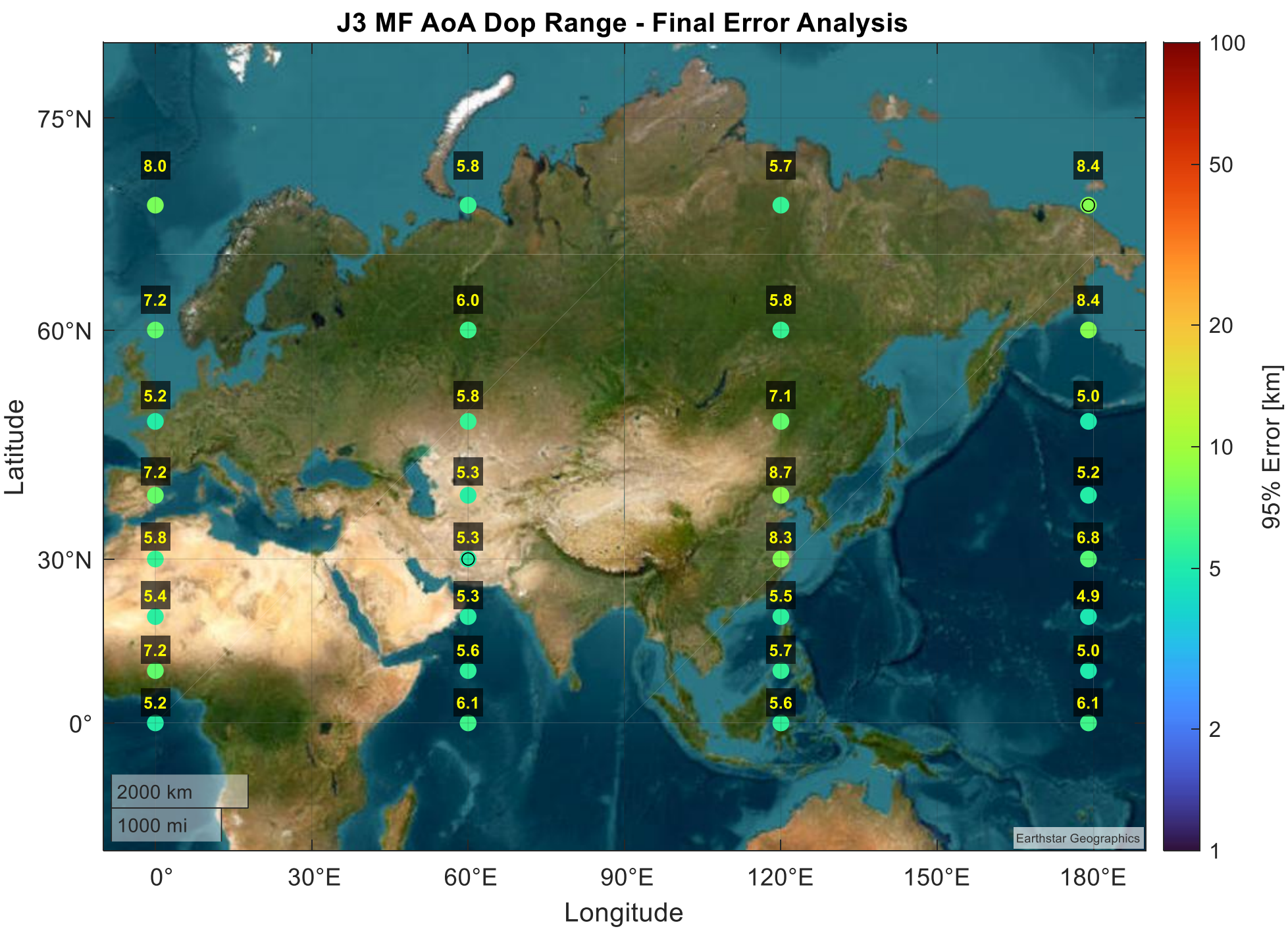

**Figure 18.** Estimation accuracy for J3 fixed locations grid for Worst Scenario

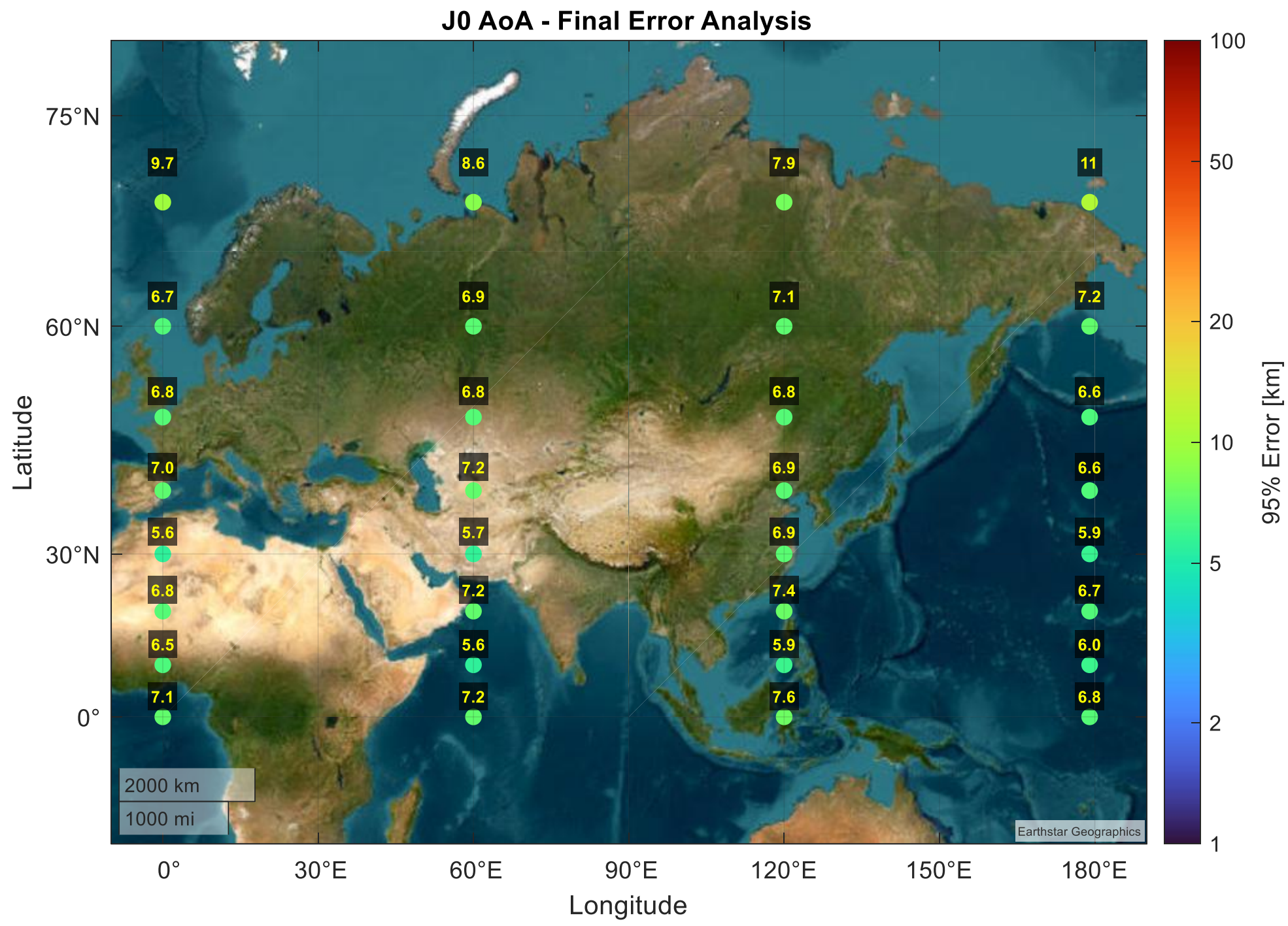

**Figure 19.** Estimation accuracy for J0 fixed locations grid for Best Scenario

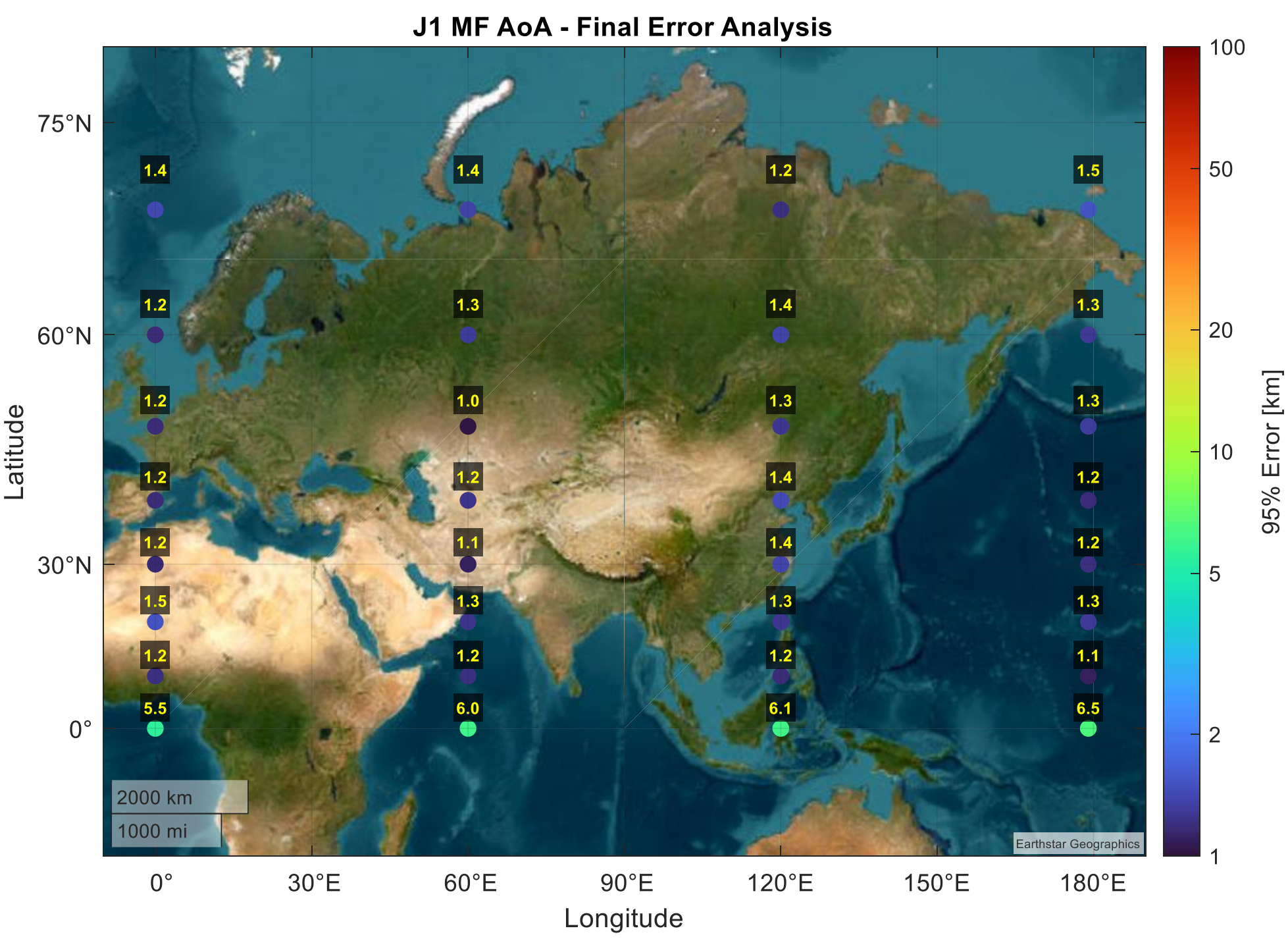

**Figure 20.** Estimation accuracy for J0 fixed locations grid for Worst Scenario

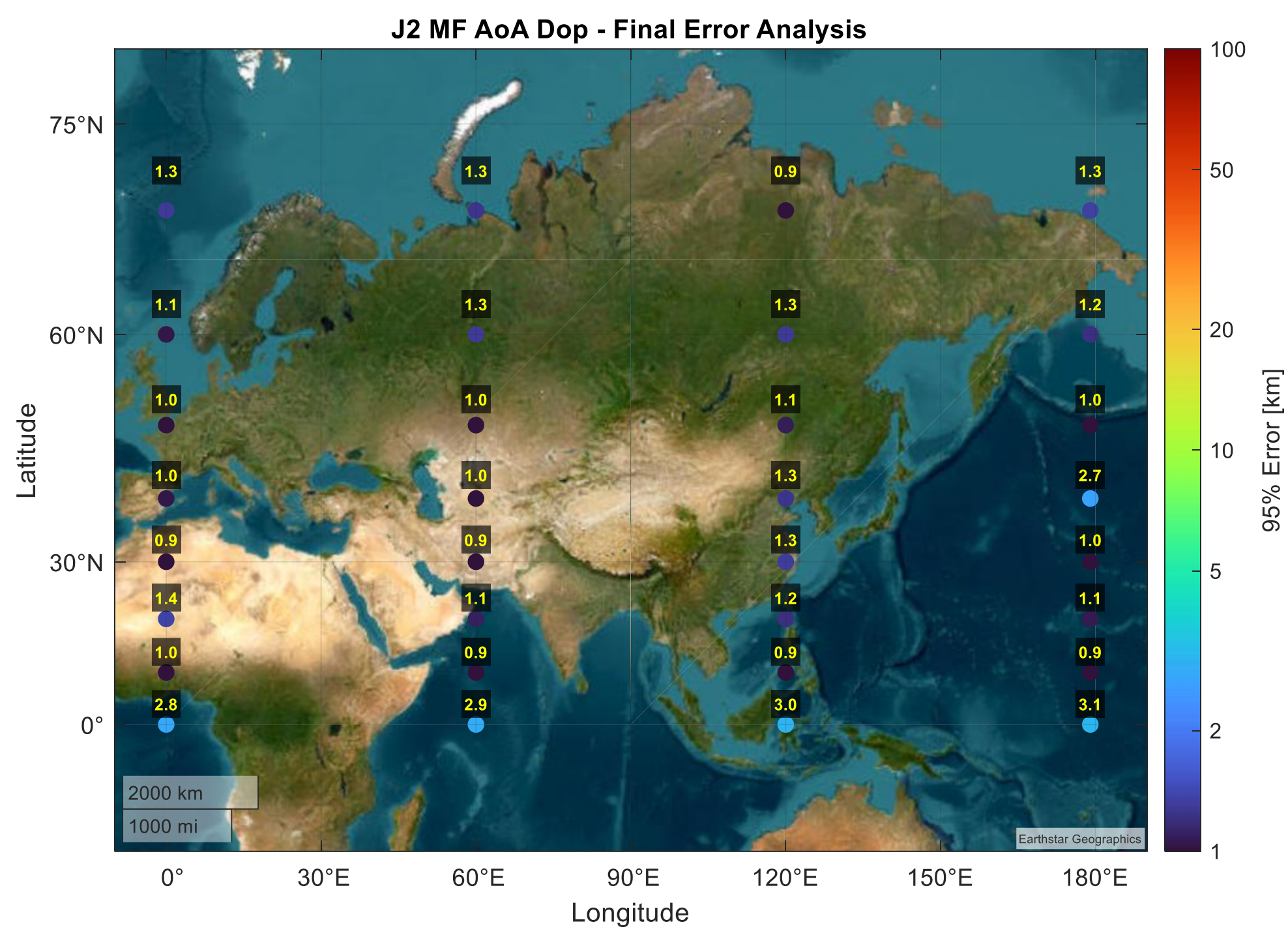

**Figure 21.** Estimation accuracy for J0 fixed locations grid for Worst Scenario

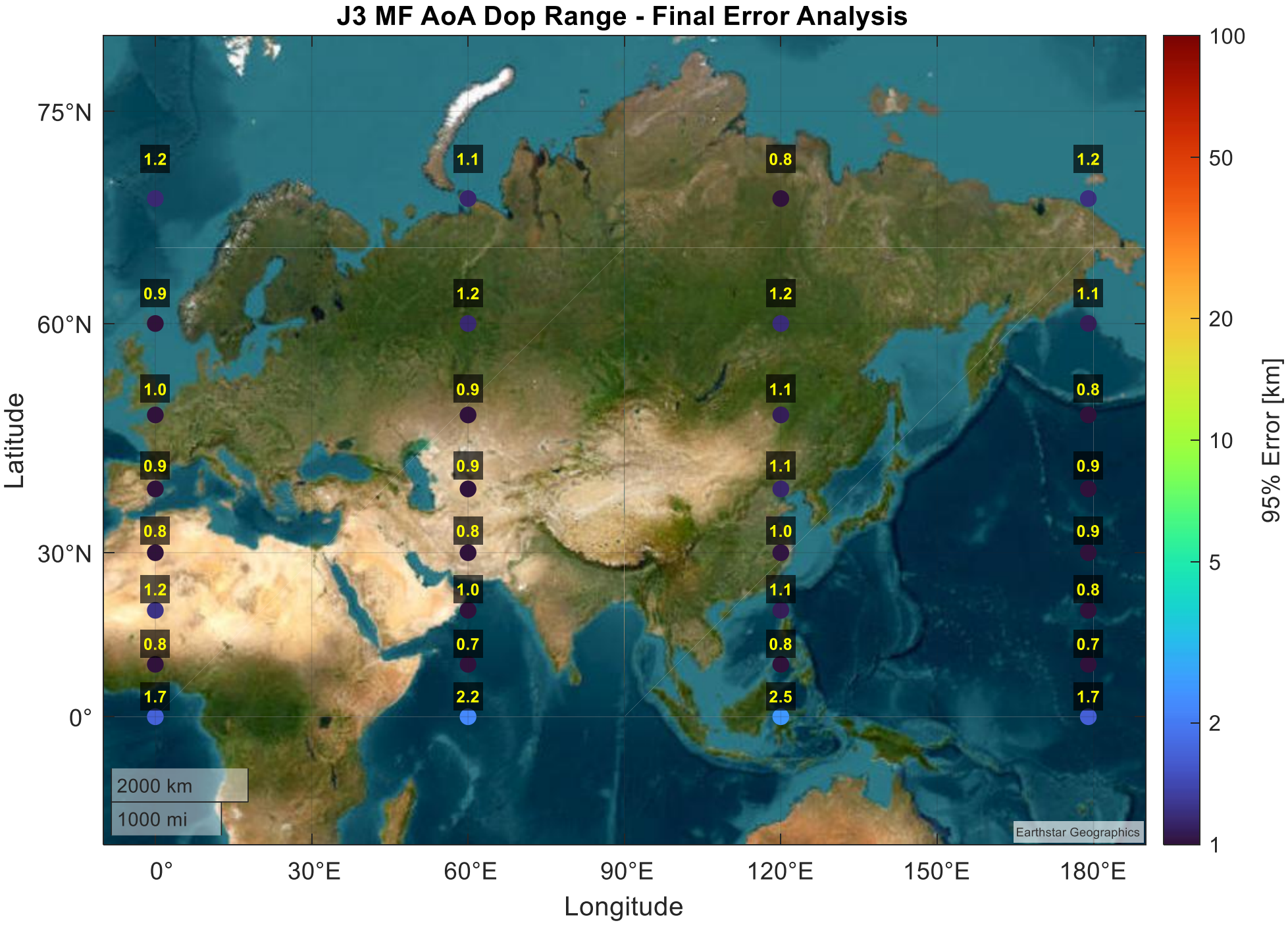

**Figure 22.** Estimation accuracy for J0 fixed locations grid for Worst Scenario

## References

1. 3GPP. (2024). *Technical Specification Group Radio Access Network; NR; NR and NG-RAN Overall Description; Stage 2 (Release 18)* (3GPP TS 38.300 version 18.1.0). https://www.3gpp.org/ftp/Specs/archive/38_series/38.300/
2. 3GPP. (2024). *Technical Specification Group Radio Access Network; NR; User Equipment (UE) radio transmission and reception; Part 5: Satellite access radio frequency (RF) and performance requirements (Release 18)* (3GPP TS 38.101-5 version 18.6.0). https://www.3gpp.org/ftp/Specs/archive/38_series/38.101-5/
3. J. Wiseman, (2024). GNSS interference map. https://gpsjam.org/.
4. Supantha Mukherjee, Cassell Bryan-Low and Parisa Hafezi, "Iranians tap Musk's Starlink to skirt internet blackout, sources say", Reuters, 13 January 2026. Available at: https://www.reuters.com/technology/iranians-tap-musks-starlink-skirt-internet-blackout-sources-say-2026-01-13/
5. 3GPP. (2025). *Technical Specification Group Radio Access Network; Revised SID: Study on GNSS (Global Navigation Satellite System) resilient NR-NTN (Non-Terrestrial Networks) operation* (3GPP TSG RAN Meeting #107 RP-251933). https://www.3gpp.org/ftp/tsg_ran/TSG_RAN/TSGR_107/Docs/
6. 3GPP. (2025). *Technical Specification Group Radio Access Network; Draft TR 38.742:* Study on GNSS (Global Navigation Satellite System) resilient NR-NTN (Non-Terrestrial Networks) operation (3GPP TSG RAN Meeting TR Draft). https://www.3gpp.org/ftp/Specs/archive/38_series/38.742/
7. 3GPP. (2026). *Technical Specification Group Radio Access Network*, R1-2602939 (3GPP TSG RAN WG1 Meeting #124 R1-2602939). https://www.3gpp.org/ftp/tsg_ran/WG1_RL1/TSGR1_124/Docs/
8. 3GPP. (2026). *Technical Specification Group Radio Access Network*; R1-2601077 (3GPP TSG RAN WG1 Meeting #124 R1-2601077). https://www.3gpp.org/ftp/tsg_ran/WG1_RL1/TSGR1_124/Docs/
9. 3GPP. (2026). *Technical Specification Group Radio Access Network*; R1-2601041 (3GPP TSG RAN WG1 Meeting #124 R1-2601041). https://www.3gpp.org/ftp/tsg_ran/WG1_RL1/TSGR1_124/Docs/
10. J. Zhu, Y. Sun, and M. Peng, "Timing advance estimation in Low Earth Orbit satellite networks," *IEEE Transactions on Vehicular Technology*, vol. 73, no. 3, pp. 4366–4382, Mar. 2024.
11. He, C., Zhang, M., & Guo, F. (2019). "Bias Compensation for AOA-Geolocation of Known Altitude Target Using Single Satellite." IEEE Access.
12. Sun, Y., & Ho, K. C. (2011). "An Accurate Closed-Form Solution for Emitter Localization Using AOA and Altitude Measurements." IEEE Transactions on Signal Processing.
13. Cho, S.-W.; Lee, J.-H. Efficient implementation of the Capon beamforming using the Levenberg-Marquardt scheme for two dimensional AOA estimation. Prog. Electromagn. Res. 2013, 137, 19–34, https://doi.org/10.2528/pier12122711.
14. 3GPP, "Study on GNSS resilient NR-NTN operation," 3rd Generation Partnership Project (3GPP), Technical Report (TR) 38.742, Rel. 20, 2025.
15. Neinavaie, M.; Khalife, J.; Kassas, Z.M. Exploiting Starlink signals for navigation: First results. In Proceedings of the 34th International Technical Meeting of the Satellite Division of the Institute of Navigation (ION GNSS+ 2021), St. Louis, MO, USA, 20–24 September 2021; pp. 2766–2773. https://doi.org/10.33012/2021.18122
16. Komodromos, Z.M.; Qin, W.; Humphreys, T.E. Signal simulator for Starlink Ku-band downlink. In Proceedings of the 36th International Technical Meeting of the Satellite Division of the Institute of Navigation (ION GNSS+ 2023), Denver, CO, USA, 11–15 September 2023; pp. 2798–2812. https://doi.org/10.33012/2023.19308
17. Lichtenberger, C.A.; Binder, F.; Picchi, O.M.; Menzione, F.; Pany, T. Using a block-processing discriminator for precise tracking of Starlink signals. In Proceedings of the 38th International Technical Meeting of the Satellite Division of the Institute of Navigation (ION GNSS+ 2025), Baltimore, MD, USA, 8–12 September 2025; pp. 3479–3491. https://doi.org/10.33012/2025.20465
18. Binder, F., Lichtenberger, C. A., Pany, T., Picchi, O. M., Winkel, J., Rotoloni, M., Menzione, F., & Soualle, F. (2026). Remarks on tracking 5G NTN signals with a GNSS-centric SDR: CRLB, GPU/CPU-based correlation and C/N0-estimation. *Proceedings of the 2026 International Technical Meeting of The Institute of Navigation*, 1–15.
19. Soualle, F. *et al.* (2025). Experimental framework for hybrid navigation terminals exploiting LEO PNT constellations. *Proceedings of the 38th International Technical Meeting of the Satellite Division of The Institute of Navigation (ION GNSS+ 2025)*, 2370–2400. https://doi.org/10.33012/2025.20328

20. 3GPP. (2019). *Solutions for NR to support non-terrestrial networks (NTN)* (Technical Report TR 38.821; Version 16.0.0). 3rd Generation Partnership Project.
21. 3GPP. Technical Specification Group Radio Access Network; Study on New Radio (NR) to Support Non-Terrestrial Networks (Release 15); Technical Report (TR) 38.811; Version 15.4.0; 3GPP: Valbonne, France, 2020.
22. Picchi, O.M.; Menzione, F.; Soualle, F.; del Peral-Rosado, J.A.; Julien, O.; Binder, F.; Majorana, A.M.; Boyero, J.P. Fused PNT performance assessment from first to second generation user terminal for an IRIS2 constellation. In Proceedings of the 38th International Technical Meeting of the Satellite Division of The Institute of Navigation (ION GNSS+ 2025), Baltimore, MD, USA, 8–12 September 2025; pp. 1305–1326. https://doi.org/10.33012/2025.20412
23. Picchi, O.M.; Menzione, F.; Soualle, F.; del Peral-Rosado, J.A. Fused PNT on future wideband European non-terrestrial network infrastructure. In Proceedings of the 2025 IEEE/ION Position, Location and Navigation Symposium (PLANS), Salt Lake City, UT, USA, 28 April–1 May 2025; pp. 417–428. https://doi.org/10.1109/PLANS61571.2025.11028398
24. A. Rozé, M. Crussière, M. Hélard and C. Langlais, "Comparison between a hybrid digital and analog beamforming system and a fully digital Massive MIMO system with adaptive beamsteering receivers in millimeter-Wave transmissions," 2016 International Symposium on Wireless Communication Systems (ISWCS), Poznan, Poland, 2016, pp. 86-91, doi: 10.1109/ISWCS.2016.7600880.
25. Stoica, P.; Nehorai, A. MUSIC, maximum likelihood, and Cramer-Rao bound. IEEE Transactions on Acoustics, Speech, and Signal Processing 1989, 37, 720–741. https://doi.org/10.1109/29.17564
26. M. Carlin, P. Rocca, G. Oliveri, F. Viani, and A. Massa, "Directions-of-Arrival Estimation Through Bayesian Compressive Sensing Strategies," IEEE Transactions on Antennas and Propagation, vol. 61, no. 7, pp. 3828–3838, Jul. 2013, doi: 10.1109/TAP.2013.2256093.3GPP. (2024). *Technical Specification Group Radio Access Network; NR; Radio Resource Control (RRC); Protocol specification (Release 18)* (3GPP TS 38.331 version 18.1.0).
27. Kaplan, E.D.; Hegarty, C.J., Eds. *Understanding GPS/GNSS: Principles and Applications*, 3rd ed.; Artech House: Norwood, MA, USA, 2017.
28. Bar-Shalom, Y.; Li, X.R.; Kirubarajan, T. Estimation with Applications to Tracking and Navigation: Theory Algorithms and Software; John Wiley & Sons: Hoboken, NJ, USA, 2004.
29. Soualle, F. *et al.* (2024). New generation of PNT user terminals exploiting hybridization with LEO constellations. *Proceedings of the 37th International Technical Meeting of the Satellite Division of The Institute of Navigation (ION GNSS+ 2024)*, 887–947. https://doi.org/10.33012/2024.19780
30. ITU-R Recommendation M.1643: Technical and operational requirements for aircraft earth stations of aeronautical mobile-satellite service operating in the band 14–14.5 GHz / 29.5–30 GHz (Ka-band).
31. 3GPP. (2024). *Technical Specification Group Radio Access Network; NR; Base Station (BS) radio transmission and reception (Release 18)* (3GPP TS 38.104 version 18.5.0). https://www.3gpp.org/ftp/Specs/archive/38_series/38.104/
32. Menzione, F. *et al.* (2024). Design and Testing of NewSpace Galileo Receiver for LEO Precise Onboard Orbit Determination in the Horizon 2020 IOV/IOD Mission. In *Proceedings of the 37th International Technical Meeting of the Satellite Division of The Institute of Navigation (ION GNSS+ 2024)*, Baltimore, MD, USA, 16–20 September 2024; pp. 1017–1036. https://doi.org/10.33012/2024.19759
33. Betz, J. W., & Kolodziejski, K. R. (2009). Generalized theory of code tracking with an early-late discriminator part I: Lower bound and coherent processing. *IEEE Transactions on Aerospace and Electronic Systems*, *45*(4), 1538–1556.
34. 3GPP. Technical Specification Group Radio Access Network; Study on New Radio (NR) to Support Non-Terrestrial Networks (Release 15); Technical Report (TR) 38.811; Version 15.4.0; 3GPP: Valbonne, France, 2020.
35. Van Sickle, J. *Basic GIS Coordinates*; CRC Press LLC: Boca Raton, FL, USA, 2004.
36. Borre, K.; Akos, D.M.; Bertelsen, N.; Rinder, P.; Jensen, S.H. *A Software-Defined GPS and Galileo Receiver: A Single-Frequency Approach*; Birkhäuser: Boston, MA, USA, 2007.
37. I. Edjekouane, A. González-Garrido, J. Querol and S. Chatzinotas, "User Equivalent Range Error and Positioning Accuracy Analysis for ToA-Based Techniques Using PRS and SSB in 5G/6G NTN," in *IEEE Open Journal of the Communications Society*, vol. 6, pp. 9052-9072, 2025, doi: 10.1109/OJCOMS.2025.362366 Kodheli, O., Astro, A., Querol, J., Gholamian, M., Kumar, S., et al. (2021). Random Access Procedure Over Non-Terrestrial Networks: From Theory to Practice. *IEEE Access*, *9*, 109130–109143. https://doi.org/10.1109/access.2021.3101291
38. 3GPP. (2024). *Technical Specification Group Radio Access Network; NR; Physical channels and modulation (Release 18)* (3GPP TS 38.211 version 18.1.0). https://www.3gpp.org/ftp/Specs/archive/38_series/38.211/

39. 3GPP. (2024). *Technical Specification Group Radio Access Network; NR; Medium Access Control (MAC) protocol specification (Release 18)* (3GPP TS 38.321 version 18.1.0). https://www.3gpp.org/ftp/Specs/archive/38_series/38.321/
40. Mohamed, S.E.Z.; *et al.*, C.D. Satellite-Based Localization of IoT Devices Using Joint Doppler and Angle-of-Arrival Estimation. *Remote Sens.* **2023**, *15*, 5603.
41. Yu, H.; *et al.*, An Efficient Closed-Form Solution for Single Satellite Geolocation Using Hybrid AOA and FDOA Measurements. IEEE Signal Process. Lett. 2020, 27, 1819–1823.
42. Xie, C.; Fang, X.; Yang, X. Improved Kalman Filtering Algorithm Based on Levenberg–Marquart Algorithm in Ultra-Wideband Indoor Positioning. Sensors 2024, 24, 7213.
43. Chen, L.; Ma, Y. A Modified Levenberg–Marquardt Method for Solving System of Nonlinear Equations. J. Appl. Math. Comput. 2023, 69, 2019–2040.
44. Piccolo, A.; *et al.*, J. JRC Testing and Demonstration Hub for the GNSS Component of the EU Space Programme: Inventory of GNSS Testing Capabilities. EUR 31883 EN, Publications Office of the European Union: Luxembourg, 2024. https://doi.org/10.2760/726940
45. Requtech AB. RESA L Electronically Scanned Antenna Terminal Datasheet. Available online: https://requtech.com/products/resa-l/ (accessed on 20 August 2026).